\documentclass[graybox]{svmult}

\usepackage{type1cm}        
\usepackage{makeidx}         
\usepackage{graphicx}        
\usepackage{multicol}       
\usepackage[bottom]{footmisc}
	
\usepackage{newtxtext}       
\usepackage{newtxmath}  

\usepackage{hyperref} 
\usepackage{url}
\usepackage{booktabs}
\usepackage{xcolor}
     
\makeindex             

\usepackage{graphicx}
\graphicspath{{./figs/}{./figsnew/}}
\usepackage{caption}
\usepackage{subcaption}

\newcommand{\R}{\mathbb{R}}
\usepackage{float}

\usepackage{algorithm}
\usepackage[noend]{algpseudocode}
\usepackage{tikz}

\newcommand {\mm}[1] {\ifmmode{#1}\else{\mbox{\(#1\)}}\fi}
\newcommand{\Xspace}        {\mm{{\mathbb X}}}
\newcommand{\Rspace}        {\mm{{\mathbb R}}}
\newcommand{\Ucal}{\mathcal{U}}
\newcommand{\Vcal}{\mathcal{V}}
\newcommand{\Wcal}{\mathcal{W}}
\newcommand{\Nrv}{\mathrm{Nrv}}
\newcommand{\St}{\mathrm{St}}

\newcommand{\LH}{\mathtt{LH}}
\newcommand{\LE}{\mathtt{LE}}

\newcommand{\LHD}{\mathtt{LHD}}
\newcommand{\LREC}{\mathtt{LREC}}
\newcommand{\LED}{\mathtt{LED}}

\newcommand{\para}[1]        {\vspace{1mm}\noindent{\textbf{#1}}}
\newcommand{\etal} {{\textit{et al.}}}
\newcommand{\ie} {{\textit{i.e.}}}

\usepackage[capitalize]{cleveref}

\usepackage{mathptmx}
\DeclareMathAlphabet{\mathcal}{OMS}{cmsy}{m}{n}
\usepackage{mathtools}
\usepackage{lmodern}
\DeclareMathDelimiter{(}{\mathopen} {operators}{"28}{largesymbols}{"00}
\DeclareMathDelimiter{)}{\mathclose}{operators}{"29}{largesymbols}{"01}

\begin{document}

\title*{Stitch Fix for Mapper and Topological Gains}

\titlerunning{Stitch Fix for Mapper and Topological Gains} 

\author{Youjia Zhou, Nathaniel Saul, Ilkin Safarli, Bala Krishnamoorthy, Bei Wang}

\institute{
Youjia Zhou \at University of Utah, Salt Lake City, UT. 
      \email{zhou325@sci.utah.edu}
\and Nathaniel Saul \at  Washington State University, Vancouver, WA. 
      \email{nat@riverasaul.com} 	     
\and Ilkin Safarli \at University of Utah, Salt Lake City, UT. 
      \email{ilkin@sci.utah.edu}
\and Bala Krishnamoorthy \at Washington State University, Vancouver, WA.
      \email{kbala@wsu.edu}
\and Bei Wang \at University of Utah, Salt Lake City, UT. 
      \email{beiwang@sci.utah.edu}}

\maketitle

\abstract{
The mapper construction is a powerful tool from topological data analysis that is designed for the analysis and visualization of multivariate data.
In this paper, we investigate a method for stitching a pair of univariate mappers together into a bivariate mapper, and study topological notions of information gains, referred to as topological gains, during such a process. 
We further provide implementations that visualize such topological gains for mapper graphs.

}


\section{Introduction}
\label{sec:introduction}

The mapper construction is one of the main tools in topological data analysis and visualization used for the study of multivariate data~\cite{SinghMemoliCarlsson2007}.
It takes as input a multivariate function defined on the data and produces a topological summary of the data using a cover of the range space of the function.
For a given cover, such a summary converts the mapping into a simplicial complex suitable for data exploration. 

In this paper, we take a \emph{constructive} viewpoint of a multivariate function
$f: \Xspace \to \Rspace^d$ defined on a topological space $\Xspace$ and consider it as a vector of continuous, real-valued functions defined on a shared domain, 
$f = (f_1, f_2, \cdots,  f_d), f_i: \Xspace \to \Rspace$, 
where each $f_i$ (referred to as a \emph{filter function}) gives rise to a univariate \emph{mapper} (or \emph{mapper construction}).
We investigate a method for stitching a pair of univariate mappers together and study topological notions of information gains, referred to as \emph{topological gains}, from such a process. 
Our notion of topological gain is loosely inspired by the concept of \emph{information gain} used in the construction of decision trees, which is computed by comparing the entropy of the dataset before and after a transformation.  
Topology gain, in our context, measures the change in topological information by comparing the homology or entropy of the mapper before and after the stitching process.
In particular, we aim to assign a measure that captures information about how each filter function contributes to the topological content of the stitched result, and how the two filter functions are topologically correlated. 

We are also inspired by the ideas of stepwise regression for model selection and scatterplot matrices for visualization.
For a set of variables $x_1, x_2, \ldots, x_d$, the \emph{stepwise regression}~\cite{Efroymson1960, Hocking1976} iteratively incorporates variables into a regression model based on some criterion.
A measure of topological gain can be used as a criterion for choosing filter functions and constructing a single ``best'' multivariate mapper.
The \emph{scatterplot matrix}~\cite{ElmqvistDragicevicFekete2008} shows all pairwise scatterplots for the set of variables on a single $d \times d$ matrix, where each scatterplot illustrates the degree of correlation between two variables.
We are interested in a topological analogue of the scatterplot matrix for a set of filter functions $f_1, f_2, \ldots, f_d$, referred to as \emph{mapper matrix}, and in studying the degree of \emph{topological correlation} between filter functions.

\para{Contributions.}
Our contributions are as follows: 
\begin{itemize}
\item We define a composition (or stitching) operation for mappers (Definition~\ref{composition}) and show its equivalence to the standard construction (Theorem~\ref{equivalence-theorem}). We then provide an algorithm for carrying out this composition (Algorithm~\ref{alg:stitch}). Although the composition produces identical results as the standard construction, we focus on interrogating the composition process itself to study and quantify structural differences between a bivariate mapper and its corresponding univariate mappers.  
\item We consider three measures that quantify topological gains during the stitching process (\autoref{sec:information-gains}). To the best of our knowledge, this is the first time information-theoretic measures are used in the study of mapper constructions. 
\item We end by describing our efforts in studying topological gains between filter functions via interactive visualization of a mapper graph matrix, using synthetic and real-world datasets. 
\item Our visualization tool is open source via GitHub at \url{https://github.com/tdavislab/mapper-stitching}.
\end{itemize}

\section{Related Work}
\label{sec:related-work}

The mapper construction can be considered as a discrete approximation~\cite{MunchWang2016} of the \emph{Reeb space}~\cite{EdelsbrunnerHarerPatel2008}, which is a topological descriptor of a multivariate function. 
The \emph{Reeb graph} is a special type of a Reeb space for a univariate function, which has been actively studied in recent years~\cite{BiasottiGiorgiSpagnuolo2008}. 
Similarly, the \emph{mapper graph} is a special type of mapper for a univariate function, approximating the Reeb graph under certain conditions~\cite{BrownBobrowskiMunch2020}.  
The mapper construction has emerged as a practical and effective tool to solve a number of problems in data science~\cite{Alagappan2012,LumSinghLehman2013, NicolauLevineCarlsson2011,RoblesHajijRosen2018,SaggarSpornsGonzalez-Castillo2018}. 
The mapper construction has a number of variants, including the \emph{$\alpha$-Reeb graph}~\cite{ChazalSun2014}, 
\emph{extended Reeb graph}~\cite{BarralBiasotti2014}, 
\emph{multi-resolutional Reeb graph}~\cite{HilagaShinagawaKohmura2001}, 
\emph{multiscale mapper}~\cite{DeyMemoliWang2016},  \emph{multinerve mapper}~\cite{CarriereOudot2018},  \emph{joint contour net}~\cite{CarrDuke2013, CarrDuke2014}, and enhanced mapper graphs~\cite{BrownBobrowskiMunch2020}; see~\cite{YanMasoodSridharamurthy2021} for a survey. 

Although the mapper construction has been widely appreciated by the practitioners, many open questions remain regarding its theoretical properties, such as its information content~\cite{CarriereOudot2018,DeyMemoliWang2017,GasparovicGommelPurvine2018}, stability~\cite{BrownBobrowskiMunch2020,CarriereOudot2018}, and convergence~\cite{Babu2013,DeyMemoliWang2017,MunchWang2016,SinghMemoliCarlsson2007}; see~\cite{BrownBobrowskiMunch2020} for a discussion. 
The mapper construction can be considered as a ``lossy compression'' of the information from the original data. 
To quantify its information content, Dey et al.~\cite{DeyMemoliWang2017} showed that the 1-dimensional homology of the mapper is no richer than the domain itself.
Carri\'{e}re and Oudot~\cite{CarriereOudot2018} quantified the information encoded in the mapper using the extended persistence diagram of its corresponding Reeb graph.
Different from previous approaches, our work quantifies the topological gain (in an information-theoretic sense) of a bivariate mapper when it is constructed by stitching a pair of univariate mappers. 

There are several open-source implementations of the mapper algorithm, including \emph{Python Mapper}~\cite{MullnerBabu2013}, \emph{KeplerMapper}~\cite{VeenSaulEargle2019,VeenSaulEargle2019b}, \emph{giotto-tda} library~\cite{TauzinLupoTunstall2020}, \emph{Gudhi}~\cite{GUDHI2020}, \emph{Mapper Interactive}~\cite{ZhouChalapathiRathore2021}, and its domain-specific adaptations~\cite{KamruzzamanKalyanaramanKrishnamoorthy2019,ZhouKamruzzamanSchnable2021}. 
In particular, \emph{Mapper interactive} provides a simple but effective strategy for speeding up mapper graph computations by precomputing the distance matrix of points within each interval using a highly optimized function within \emph{scikit-learn}~\cite{ZhouChalapathiRathore2021}; it also comes with a GPU implementation.

Hajij~\etal~\cite{HajijAssiriRosen2020} studied the computation of mapper in parallel.  
The main idea is to decompose the computation onto a set of processors by decomposing the space of interest into 
multiple smaller, partially overlapping subspaces. 
Each subspace is processed independently by a processing unit, resulting in a mapper construction on the subspace. 
The final mapper graph on the entire space is obtained by merging together the individual pieces gathered from subspaces.

In comparison with the work of Hajij~\etal~\cite{HajijAssiriRosen2020}, we focus on a completely different problem. 
Their approach \cite{HajijAssiriRosen2020} produces a final mapper graph by stitching together mapper graphs constructed  from spatially distributed individual subspaces; while our Algorithm~\ref{alg:stitch}, in the univariate setting, produces a final mapper graph by stitching together mapper graphs constructed from individual filter functions.  
Specifically, each filter function gives rise to a univariate mapper. 
We stitch two univariate mappers into a bivariate mapper and study the topological gains from the stitching process using information theory.  
In the univariate setting, although both approaches produce the same final mapper graph, our work is not about the efficient computation of the mapper construction, but rather, we care about the stitching process itself and how much information is gained during the stitching process, moving from a univariate mapper to a multivariate mapper.

\section{Main Theoretical Result}
\label{sec:prelim}

We assume the readers are familiar with concepts in algebraic and computational topology such as homology (see the book by Edelsbrunner and Harer~\cite{EdelsbrunnerHarer2010} for an introduction).  
Given a space $\Xspace$, a function $f: \Xspace \to \Rspace^d$, and a cover $\Ucal = \{U_i\}$ of $f(\Xspace)$, we define the pullback cover $f^*(\Ucal)$ of $\Xspace$ as the cover obtained by decomposing each $f^{-1}(U_i)$ into its path-connected components $\cup_{j=1}^{k_i} u_{ij}$.  
The \emph{mapper}~\cite{SinghMemoliCarlsson2007} is then a simplicial complex defined as the nerve of this pullback cover $ M(f, \Ucal) := \Nrv(f^*(\Ucal))$.

\begin{figure}[!ht]
\centering
\includegraphics[width=0.58\columnwidth]{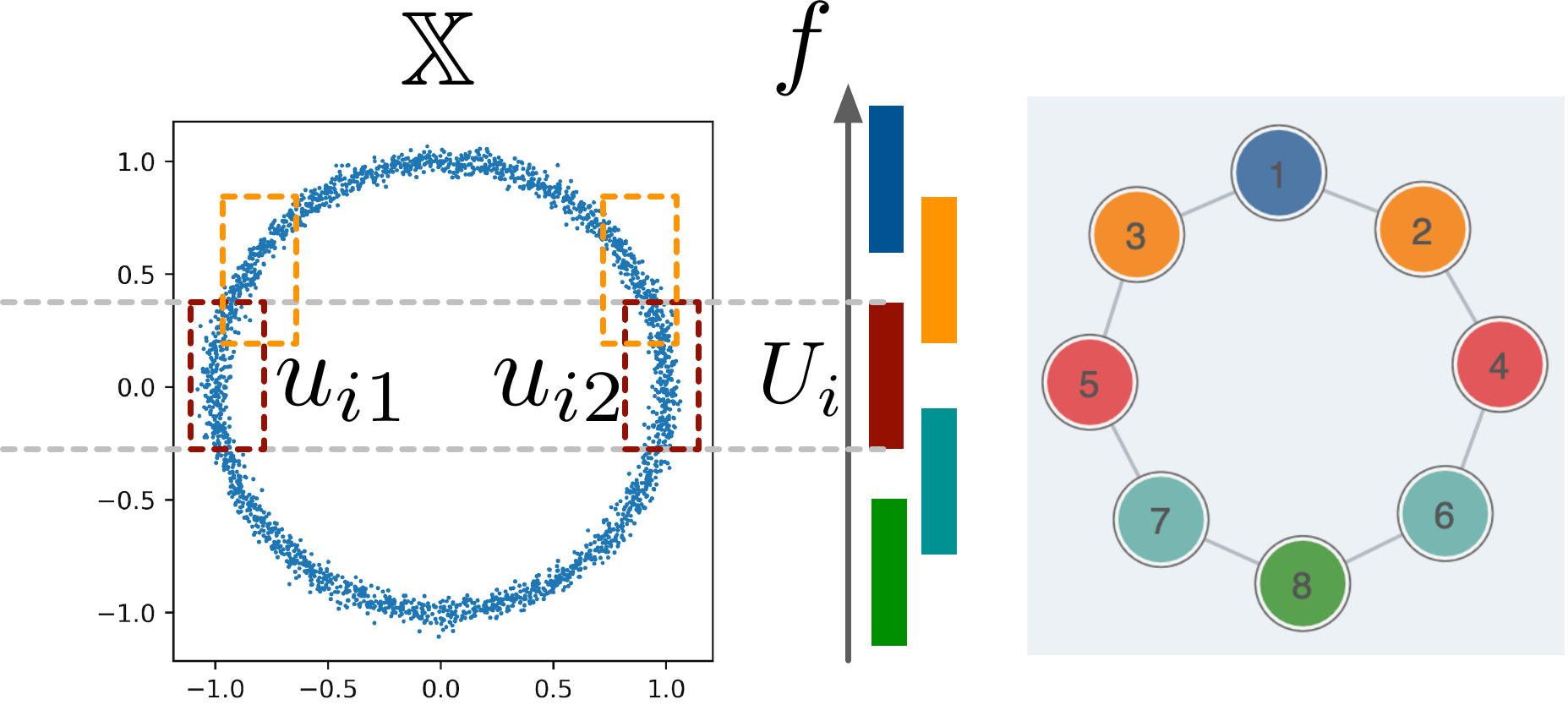}
\vspace{-2mm}
\caption{The mapper of a height function $f$ defined on a 2-dimensional  point cloud sample $\Xspace$ of a circle. Mapper parameters: $n = 5$, $p = 33\%$.}
\label{fig:circle-mapper}
\vspace{-2mm}
\end{figure}

For simplicity, we describe the mapper construction by assuming $\Xspace$ to be a point cloud equipped with a univariate function $f:\Xspace \to \Rspace$.
Several parameters are relevant to the construction of its mapper. 
First is the number of intervals (of uniform lengths) in the cover $\Ucal$ of $f(\Xspace)$, denoted as $n$ and referred to as the \emph{resolution}, giving the cover $\Ucal = \{U_1, \dots, U_n\}$.  
Second is the amount of overlap $p$ between the intervals in $\Ucal$ (e.g., $20\%$ overlap). 
Finally, there are parameters associated with a clustering algorithm (e.g., DBSCAN~\cite{EsterKriegelSander1996}) that clusters points in $f^{-1}(U_i)$ into connected components. 
These clusters of points form a pullback cover of $\Xspace$, and the mapper is the nerve of such a cover. 
An example of a univariate mapper is shown in~\autoref{fig:circle-mapper} for a height function $f:\Xspace \to \Rspace$ defined on a 2-dimensional  point cloud sample of a circle, for $n = 5$ and $p = 33\%$. 
Note that the inverse $f^{-1}(U_i)$ of the red interval $U_i$ is decomposed into two clusters $u_{i1}$ and $u_{i2}$, forming part of the pullback cover of $\Xspace$. 

On the other hand, if $f$ becomes a bivariate function,  $f = (f_1, f_2)$ for $ f_i : \Xspace \to \Rspace$ ($i = 1, 2$), then the cover of $f(\Xspace)$ consists of rectangles and the resulting mapper is referred to as a  bivariate mapper. 

\begin{definition}[Composition]
\label{composition}
Given $f, g: \Xspace \to \Rspace$ and covers $\Ucal = \{U_i\}$ and $\Vcal=\{V_j\}$ of their respective images $f(\Xspace)$ and $g(\Xspace)$, we construct a composed cover $\Wcal$ of $\Xspace$ from $f^*(\Ucal)$ and $g^*(\Vcal)$ by taking the connected components of the set $\{ U' \cap V' \mid \forall U' \in f^*(\Ucal), \forall V' \in g^*(\Vcal),  U' \cap V' \ne \emptyset\}$, where $U' \in f^*(\Ucal)$ and $V' \in g^*(\Vcal)$ are path-connected cover elements of $\Xspace$: 
\begin{align*}
\Wcal  = & \{ W_k \mid \cup_k W_k = U' \cap V', \forall U' \in f^*(\Ucal), \forall V' \in g^*(\Vcal),  U' \cap V' \ne \emptyset,\\ 
& ~~ W_k \text{ is path-connected} \}.
\end{align*}
We define the composed mapper as the nerve of this cover $\Wcal$, 
$$ S(M(f,\Ucal), M(g,\Vcal)) := \Nrv(\Wcal).$$
\end{definition}
Under certain assumptions, this composition $S$ is equivalent to the traditional method of constructing mappers from a pair of filter functions, as described by~\autoref{equivalence-theorem}.

\begin{theorem}
\label{equivalence-theorem}
If $f$ and $g$ are continuous real-valued functions, $U_i$, $V_j$, and $U_i \times V_j$ are simply connected for all $i,j$, then
$S(M(f,\Ucal), M(g,\Vcal)) = M((f,g),\Ucal \times \Vcal)$,
the bivariate mapper constructed in the traditional manner.
\end{theorem}

\noindent\textbf{\textit{Proof sketch.}}
The proof follows directly from properties of continuous functions and connected sets. We provide a sketch here. Starting with the two covers associated with the two univariate mappers, $\Ucal$ for $f$ and $\Vcal$ for $g$, we can show that the defined set $\Wcal$ and the cover obtained from the traditional mapper construction are equivalent. Taking the nerve of each, we conclude that the resulting mappers are equivalent as well. See~\autoref{sec:proof} for details. 

Furthermore, we give an algorithm (Algorithm~\ref{alg:stitch}) that illustrates how the composition can be considerably simplified by directly incorporating simplex information from each of the two input mappers.
The algorithm that combines (or stitches) two mappers together works by tracking vertices (i.e., representatives of the path-connected pull back cover elements as a result of the $\Nrv$ operation) of the first mapper in a breadth first search fashion and combining them with vertices of the second mapper. The simplices in both mappers provide hints about which possible simplices could be in the composition. Using this information to avoid many explicit intersection checks, we
can considerably simplify and speed up the composition process. 
Although some simplices from each univariate mapper can be added directly to the composition (the \texttt{STITCH} step), others require explicitly checking the nerve condition in the mapper construction (the \texttt{FIX} step). 

We recall the relevant notation used in the following sections. 
Given $f, g: \Xspace \to \Rspace$, let $\Ucal = \{U_i\}$ and $\Vcal=\{V_j\}$ denote the cover of their images $f(\Xspace)$ and $g(\Xspace)$, respectively. 
Let $\{U'\}$ and $\{V'\}$ denote the sets of path-connected cover elements of $\Xspace$ in the pullback covers, $f^*(\Ucal)$ and $f^*(\Vcal)$, respectively. 
Let $\{u\}$ and $\{v\}$ represent the set of vertices in the mappers $M(f, \Ucal)$ and $M(g, \Vcal)$, respectively. 
Let $\Wcal = \{W\}$ denote the composed cover of $\Xspace$, and $\{w\}$ the set of vertices for the composed cover, that is, $w = \Nrv(W)$.

\section{Algorithm}
\label{app:algorithm}

The stitching (composition) algorithm, as shown in Algorithm \autoref{alg:stitch}, has two main phases, \texttt{STITCH} for each cover element and \texttt{FIX} across cover elements.
The \texttt{STITCH} phase takes all vertices from the first mapper in each cover element and stitches them together with all vertices in the second mapper.
All simplices within the cover element of the second mapper will be inherited in the composition.
The second phase \texttt{FIX} addresses simplices that cross between cover elements and uses an auxiliary procedure \texttt{COMPLETE} to construct simplices that cannot be derived directly from simplices in either of the two input mappers. 

Suppose we have two filter functions, $f, g: \Xspace \to \Rspace$ together with covers of their images, $\Ucal = \{U_i\}$ and $\Vcal=\{V_j\}$.  
We define a function, $\mu(\cdot)$ that takes a vertex $v$ in $M(f, \Ucal):= \Nrv(f^*(\Ucal))$ and returns a path-connected cover element of $\Xspace$ to which the vertex belongs.  
For a cover element $U_i \in \Ucal$ (referred to as an \emph{interval set} of $f(\Xspace)$), we consider a vertex $v \in M(f, \Ucal)$ to be in the interval set $U_i$ if $f(\mu(v)) \subseteq U_i$.
We say a simplex $\sigma$ \emph{satisfies the nerve condition} if $\cap_i \mu(v_i) \ne \emptyset$ for all $v_i \subset \sigma$. 

  For each cover element $U_i \in \Ucal$, the algorithm iterates over each path-connected component of $f^*(U_i) \cap g^*(\Vcal)$.
  Hence there exists a unique cover element $W_{jh}$ of $\Xspace$ in $\{W\}$ in Line \ref{alg:line:onetoone} of \texttt{STITCH}  for $v_j$ corresponding to the $h$-th component (for some $h$) in $\mu(v_j) \cap f^*(U_i)$.
  Also, vertex $u$ in Line \ref{alg:fix:unique} of \texttt{FIX} is unique.

\begin{figure}[ht!]
\includegraphics[width=\columnwidth]{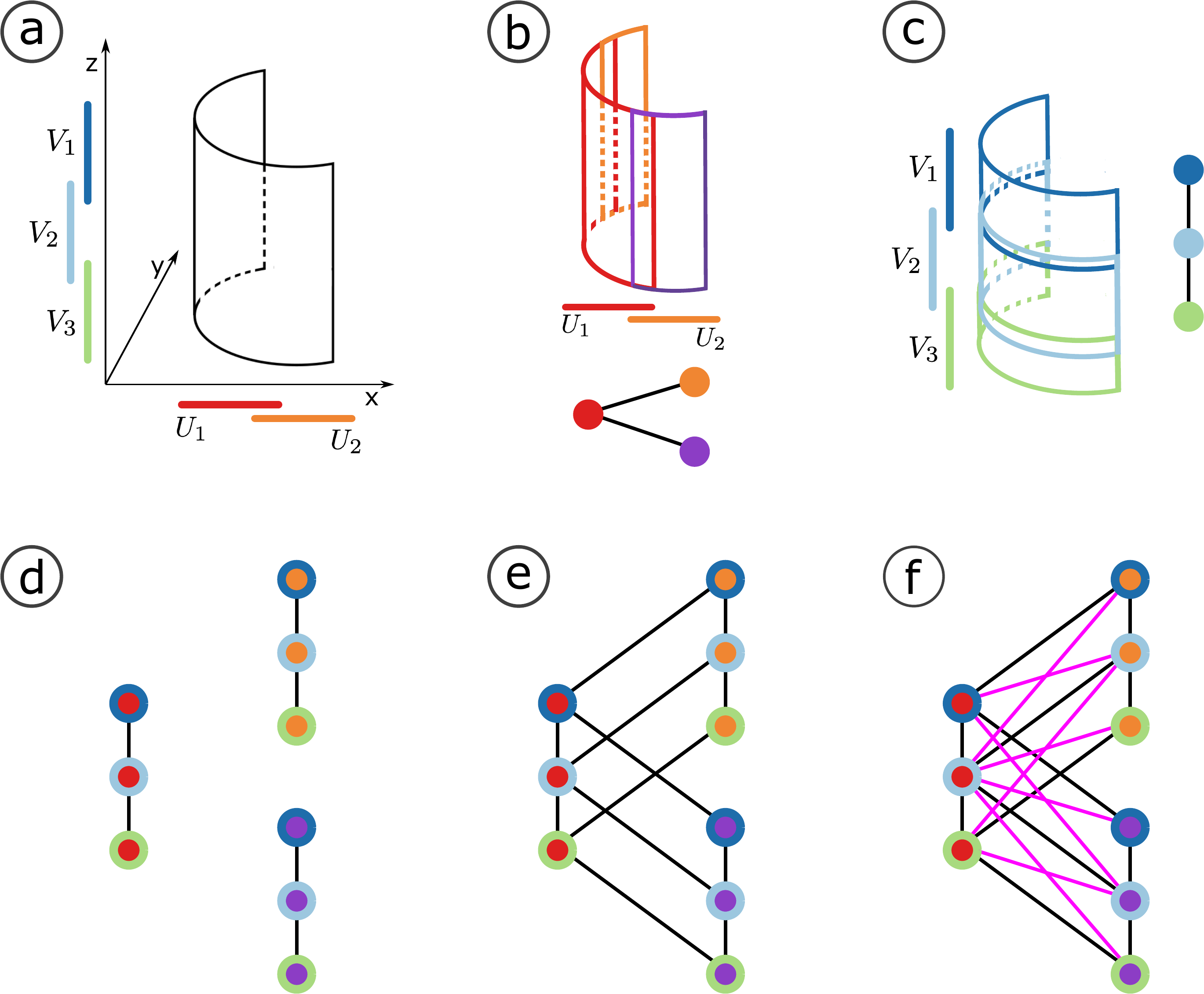}
\caption{Illustration of Algorithm \ref{alg:stitch}.
    (a) The space $\Xspace$ is a half-cylinder in $\R^3$ with height along the $z$-axis.
    We consider $f=x$ and $g=z$.
    (b) Cover $\Ucal = \{U_1, U_2\}$ of $f(\Xspace)$, path-connected pullback cover elements shown in red (for $U_1$), and in orange and purple (for $U_2$), and the mapper $M(f,\Ucal)$.
    (c) Cover $\Vcal = \{V_1, V_2, V_3\}$ of $g(\Xspace)$, corresponding path-connected pullback cover elements (in dark blue, sky blue, green), and the mapper $M(g,\Vcal)$.
    The composite cover $\Wcal$ consists of 9 path-connected cover elements determined by the intersection of each element of $f^*(\Ucal)$ (in red, orange, and purple) with each element of $g^*(\Vcal)$ (dark blue, sky blue, and green).
    The corresponding vertices in the composite mapper are shown with these pairs of colors.
    (d) The mapper composition $M$ after the \texttt{STITCH} phase of Algorithm \ref{alg:stitch}.
    (e) $M$ after the first stage of \texttt{FIX} phase (before running \texttt{COMPLETE}).
    $M$ is still $1$-dimensional at this stage.
    (f) The final composite mapper $S(M(f,\Ucal),M(g,\Vcal))$ after the \texttt{COMPLETE} phase, which adds the four tetrahedra (indicated by the pink diagonals) after adding the corresponding triangles which are their faces. 
}
\label{fig:algo}
\end{figure}

In Algorithm \ref{alg:stitch}, the set $\{u\}$ and $\{v\}$ contain vertices whereas the set $\{W\}$ contains path-connected cover elements of $\Xspace$ (we use vertex $w$ for $\Nrv(W)$). 
We make use of the notion of $p$-completion in \texttt{COMPLETE}. 
For a $p$-dimensional simplicial complex $\Sigma$, its \emph{$p$-dimensional completion}~\cite{GundertSzedlak2015} is defined to be: 
$$\Sigma \cup \left\{ \tau \in \binom{\Sigma^{(0)}}{p\text{+}1} \Big\lvert \, (\tau \setminus v) \in \Sigma,  \forall v \in \tau \right\}, $$
where $\Sigma^{(0)}$ is the vertex set of $\Sigma$. 
In our use, we reduce the set to include only simplices that satisfy the nerve condition. 
We illustrate the steps of Algorithm \ref{alg:stitch} on a simple space $\Xspace$ in \autoref{fig:algo}.

\begin{algorithm}[ht!]
\caption{Mapper Composition}
\label{alg:stitch}
\begin{algorithmic}[1]
\Require $M(f, \Ucal),~ M(g, \Vcal)$
\Ensure $S(M(f, \Ucal), M(g,\Vcal))$
\State $M \leftarrow \{\}$ \Comment{An empty simplicial complex}
\State $\Wcal \leftarrow \{\}$ \Comment{Composed cover of \Xspace; empty at start}
\For{$U_i \in \Ucal$}
  \State $\{u\} \leftarrow \{\text{vertex } u \in M(f, \Ucal) \mid f(\mu(u)) \subseteq U_i \}$
  \State $\{v\} \leftarrow \{\text{vertex } v \in M(g, \Vcal) \mid f(\mu(v)) \cap U_i \neq \emptyset \}$
  \State $\{W\} \leftarrow \{W_k \mid  \cup_{k} W_k = \mu(u_i) \cap \mu(v_j) , \text{ for } \  u_i \in \{u\}, v_j \in \{v\}, W_k \text{ is path-connected}  \}$
  \State $M \leftarrow$ \texttt{STITCH}($M$, $\{ v\}$, $\{ W\}$)
  \State $\Wcal \leftarrow \Wcal \cup \{W\}$
\EndFor
\State $M \leftarrow$ \texttt{FIX}($M$, $\Wcal$)
\State \Return $M$
\end{algorithmic}

\begin{algorithmic}[1]
  \Procedure{\texttt{STITCH}}{$M$, $\{v\}$, $\{W\}$} \Comment{Add vertices and within-interval simplices}
  \State $K \leftarrow$ subcomplex of $M(g, \Vcal)$ induced by $\{v\}$
  \For{$j \leftarrow 1, \dots, |\{v\}|$}
      \For{$h \leftarrow 1, 2, \ldots, l_j$} \Comment{Repeat for each of the $l_j$  components of $\mu(v_j) \cap f^*(U_i)$}
  	  \State $K \leftarrow (K \setminus v_j) \cup w_{jh}$ \label{alg:line:onetoone} \Comment{Replace $v_j$ with $w_{jh} = \Nrv(W_{jh})$}
      \EndFor
  \EndFor
  \State $M \leftarrow$ $M \cup K$ 
  \State \Return $M$
  \EndProcedure
\end{algorithmic}

\begin{algorithmic}[1]
  \Procedure{\texttt{FIX}}{$M$, $\Wcal$} \Comment{Add cross-interval simplices}
  \For{$W \in \Wcal$}
  	\State $u \leftarrow  \text{vertex } u \in M(f, \Ucal)$ where $W \subseteq \mu(u)$\label{alg:fix:unique}.
    \State $\Sigma \leftarrow \{\}$
    \For{$\sigma \in \{\sigma \mid u \subset \sigma, \text{simplex } \sigma \in M(f, \Ucal)\}$}
      \State $\sigma_w \leftarrow (\sigma \setminus u) \cup w$ \Comment{Replace $u$ with $w=\Nrv(W)$ in the new simplex}
      \If{$\sigma_w$ satisfies the nerve condition of $\sigma$ as before}
        \State $\Sigma \leftarrow$ $\Sigma \cup \sigma_w$
      \EndIf
    \EndFor
    \State $M \leftarrow M \cup \Sigma$ 
  \EndFor
  \State $M \leftarrow$ \texttt{COMPLETE}$(M)$
  \State \Return $M$
  \EndProcedure
\end{algorithmic}

\begin{algorithmic}[1]
  \Procedure{\texttt{COMPLETE}}{$\Sigma$} \Comment{Add higher order simplices} 
  \State $i \leftarrow 2$ 
  \While{$i \leq \dim(\Sigma)$+$1$}
  \State $\Sigma' \leftarrow \left\{\tau \in {\Sigma^{(0)} \choose i\text{+}1} \big\lvert (\tau \setminus v) \in \Sigma,  \forall v \in \tau, \tau \text{ satisfies the nerve condition} \right\}$
  \State $\Sigma \leftarrow \Sigma \cup \Sigma'$
  \State $i \leftarrow i\text{+}1$
  \EndWhile
  \State \Return $\Sigma$
  \EndProcedure
\end{algorithmic}
\end{algorithm}

Asymptotically, Algorithm \ref{alg:stitch} does not improve the runtime in comparison with the traditional algorithm in computing a bivariate mapper.
However, it provides a different perspective in constructing a bivariate mapper from stitching together a pair of univariate mappers. 
Understanding such a stitching process gives a detailed view of structural differences between the bivariate mapper and the univariate mappers. 

\vspace*{-0.1in}

\section{Proof of Theorem~\ref{equivalence-theorem}}
\label{sec:proof}

The following proof for Theorem~\ref{equivalence-theorem} shows that the composition of two univariate mappers is equivalent to the mapper directly constructed from a bivariate function encoding both filter functions.
The proof follows directly from properties of continuous functions and connected sets.

\begin{proof}
Given a pair of filter functions $f, g: \Xspace \to \Rspace$ equipped with covers of their images $\Ucal=\{U_i\}$ and $\Vcal=\{V_j\}$, respectively,  
we define $h = (f,g):\Xspace \to \Rspace^2$. The pullback cover $\overline{\Wcal} = h^*( \Ucal \times \Vcal )$ is constructed from a cover of the image $h(\Xspace)$ in the traditional manner. That is, the nerve of $\overline{\Wcal}$ gives the traditional bivariate mapper.

Recall in Definition \ref{composition} that $\Wcal$ is the path-connected components of the set $\{ U' \cap V' \mid \forall U' \in f^*(\Ucal), \forall V' \in g^*(\Vcal),  U' \cap V' \ne \emptyset \}$.
This proof will show that $\Wcal$ is equivalent to $\overline{\Wcal}$. 

First, we show that $\overline{\Wcal} \subseteq \Wcal$. 
Let $W \in \overline{\Wcal}$ be any path-connected cover element of $\Xspace$.  
  By definition of the mapper and the refinement, we have $W \subseteq \overline{W_i}$ for some 
  $\overline{W_i} = h^{-1}(U_i \times V_j)$, $U_i \in \Ucal$ and $V_j \in \Vcal$. 
  Thus, $h(W) \subseteq U_i \times V_j$.
  Note that  $\overline{W_i}$ is not necessarily path-connected. 
  
$h(W)$ can be further projected along the two filter functions such that $f(W) \subseteq U_i$ and $g(W) \subseteq V_j$.  
  Hence, $W \subseteq f^{-1}(U_i)$ and $W \subseteq g^{-1}(V_j)$. 
  Therefore, $W \subseteq f^{-1}(U_i) \cap g^{-1}(V_j)$. 
  Since $W$ is path-connected, it must be the intersection of path-connected components from $f^{-1}(U_i)$ and $g^{-1}(V_j)$, respectively. 
  That is, we must have $W = U' \cap V'$ for some path-connected components $U' \in f^{-1}(U_i)$ and $V' \in g^{-1}(V_j)$. 
  Therefore, $W \in \Wcal$; thus $\overline{\Wcal} \subseteq \Wcal$.

Second, we consider the cover $\Wcal$ of $\Xspace$ used to construct $S(M(f, \Ucal), M(g, \Vcal))$.
  We show that  $\Wcal \subseteq \overline{\Wcal}$. 
  Take a path-connected element $W \in \Wcal$.
  By definition, $W \subseteq U' \cap V'$ for some path-connected $U' \in f^*(\Ucal)$ and $V' \in g^*(\Vcal)$. 
  Thus, we have $W \subseteq U' \in f^{-1}(U_i)$ for some $U_i \in \Ucal$ and similarly $W \subseteq V' \in g^{-1}(V_j)$ for some $V_j \in \Vcal$.
  Thus, we have $f(W) \subseteq U_i$ and $g(W) \subseteq V_j$.
  It follows that $h(W) \subseteq U_i \times V_j$, and therefore, $W \subseteq h^{-1}(U_i \times V_j)$.

  This observation shows that $\overline{\Wcal}$ and $\Wcal$ are equivalent.
  Since both constructions of the mapper derive from the same cover, the resulting mappers must be equivalent.
\end{proof}

\section{Quantifying Topological Gains}
\label{sec:information-gains}

Theorem~\ref{equivalence-theorem} and Algorithm \ref{alg:stitch} inspire us to think about ways to quantify structural differences between a bivariate mapper and its corresponding univariate mappers. 
Moving from theory to practice, we consider three measures that quantify topological gains during the stitching process using homology or entropy. 
These measures are straightforward to compute, and use only information within each interval set from the \texttt{STITCH} phase of the composition algorithm.
We aim to give simple quantitative measures describing the change in information content from a univariate mapper construction to a bivariate mapper construction. 
Although these measures do not capture the complete connectivity information across interval sets, they provide a quantitative assessment of the stitching process both globally and locally. 

\para{Notations.}
Before introducing these measures, we first introduce a few notations regarding mappers and mapper graphs restricted to interval sets. 
Given mappers $K_f : = M(f,\Ucal)$ and $K_{(f,g)}:= M((f,g), \Ucal \times \Vcal)$, let 
$K_{f}(U_i)$ be the subcomplex of $K_f$ restricted to the interval $U_i \in \Ucal$, and similarly let  $K_{(f,g)}(U_i)$ be the subcomplex of $K_{(f,g)}$ restricted to the interval $U_i$.

We consider two types of restrictions. 
We construct \emph{interior subcomplexes} $\mathring{K}_{f}(U_i)$ by considering the subcomplexes of $K_f$ induced by vertices $x \in K_f$ whose function values $f(x)$ fall in the interval $U_i$.
We also construct \emph{boundary subcomplexes} $\overline{K}_{f}(U_i)$ by considering all $\sigma \in \mathring{K}_{f}(U_i)$ and their cofaces. 
Recall the \emph{star} of a simplex in a simplicial complex $K$ consists of all its cofaces in $K$, $\St(\sigma) = \{\tau \in K \mid \sigma \leq \tau\}$.  
Then we have $\overline{K}_{f}(U_i) = \{\St(\sigma)  \mid \sigma \in \mathring{K}_{f}(U_i) \}$.   
Similarly, we define $\mathring{K}_{(f,g)}(U_i)$ and $\overline{K}_{(f,g)}(U_i)$. 
If $K_f$ and $K_{(f,g)}$ are replaced by their 1-dimensional skeletons  (referred to as their \emph{mapper graphs}) denoted as $G_f$ and $G_{(f,g)}$, respectively,  then we speak of \emph{interior subgraphs} $\mathring{G}_f(U_i)$ and \emph{boundary subgraphs} $\overline{G}_f(U_i)$ accordingly. 

Both localized homological difference (\autoref{sec:LHD}) and local relative Euler characteristics (\autoref{sec:LREC}) are applicable to the mapper subcomplexes and mapper subgraphs, whereas the localized entropy differences (\autoref{sec:LED}) are applicable only to the mapper subgraphs. 

\subsection{Localized Homological Difference}
\label{sec:LHD}

The localized homological difference ($\LHD$) compares the Betti numbers for each interval set between the two mapper constructions. 
This measure bears a weak resemblance to the approach taken by Edelsbrunner \etal~\cite{EdelsbrunnerHarerNatarajan2004}, where local and global comparison measures are introduced for a pair of real-valued functions defined on a common domain; in particular, such measures can be related to the set of critical points from one function to the level sets of the other. 

Intuitively speaking, the $\LHD$ characterizes the homological information gain during the composition (stitching) process. 
Starting with a mapper $K_f$ associated with the first function $f$, $\LHD$ studies what happens within each interval set while stitching a mapper $K_g$ associated with the second function $g$. 
Let $\beta_p(K)$ denote the $p$-th Betti number of a simplicial complex $K$.

\begin{definition}[Localized Homological Difference]
Let $K_f:=M(f, \Ucal)$ be the first mapper and $K_g:=M(g, \Vcal)$ the second. 
We define $\LHD_p$ as a vector that quantifies the amount of $p$-dimensional homological information gained by stitching the second mapper onto the first one within each interval set $U_i \in \Ucal$. 
That is, we define localized homology (LH) vectors $\beta_p^f$ and $\beta_p^{(f,g)}$ to encode homological information associated with each mapper subcomplex and compute their difference (suppose $|\Ucal| = k$), 
\begin{align}
& \beta_p^f  = \left(\beta_p(K_f(U_1), \beta_p(K_f(U_2), \dots, \beta_p(K_f(U_k)) \right),\\
& \beta_p^{(f,g)}  = \left(\beta_p(K_{(f, g)}(U_1)),  \beta_p(K_{(f, g)}(U_2)), \dots,  \beta_p(K_{(f, g)}(U_k)) \right),\\
& \LHD_p\left(K_f, K_{(f,g)}\right) =   \beta_p^{(f,g)} - \beta_p^f.
\end{align}
\end{definition}
Here, $K_f(U_i)$ and $K_{(f,g)}(U_i)$ represent either interior or boundary  subcomplexes. 

Example~\ref{lhd} below demonstrates the $\LHD$ calculation for $p=1$ on a cylinder, using interior subcomplexes. 
Consider a cylinder embedded in a 3-dimensional space  centered on the origin with a circle along the $x$-$y$ plane and the tube rising along the $z$-axis.
Suppose we have three filter functions that represent the projection along the $x$, $y$, and $z$ axes, respectively.
For simplicity, let us denote these filter functions as $x$, $y$, and $z$.
The cover of each filter function is made of three equal length intervals with small overlaps spanning the range.
Clearly, the images of filter functions $x$ and $y$ are nearly identical whereas that of $z$ is distinct.
Let $K_x := M(x, \Ucal)$ and $K_z : = M(z, \Vcal)$ denote the univariate  mappers associated with filter functions $x$ and $z$, respectively. 
Also let $K_{(x,z)}:=M((x,z), \Ucal \times \Vcal)$ be a bivariate mapper.  

\begin{example}[$\LHD_1$ on a cylinder]
\label{lhd}
$$\LHD_1\left(K_x, K_{(x,z)}\right) = \begin{pmatrix}0 - 0 \\ 0-0 \\ 0-0 \end{pmatrix} =  \begin{pmatrix} 0 \\ 0 \\ 0 \end{pmatrix}$$
$$\LHD_1\left(K_z, K_{(x,z)})\right) = \begin{pmatrix}1 - 0 \\ 1-0 \\ 1-0 \end{pmatrix} =  \begin{pmatrix} 1 \\ 1 \\ 1 \end{pmatrix}$$

To illustrate this example, consider the first interval set of the $x$ projection. $K_x$ has 1 vertex in $U_1 \in \Ucal$. 
$K_{(x, z)}$ restricted to $U_1$ then consists of a line of 3 vertices with 2 edges. Thus, for the first entry in $\LHD_1$, we have
$$\beta_1\left(K_{(x, z)}(U_1)) - \beta_1(K_{x}(U_1)\right) = 0 - 0 = 0.$$ 
In contrast, the first interval set of $K_z$ contains 1 vertex, whereas $K_{(x, z)}$ within the same interval set contains a loop. Thus, for the first interval $V_1 \in \Vcal$, 
$$\beta_1\left(K_{(x, z)}(V_1)) - \beta_1(K_{z}(V_1)\right) = 1 - 0 = 1.$$
This example shows that more homological information is gained with respect to the first  homology class by stitching $K_x$ (\ie,  the mapper of filter function $x$) to $K_z$ than by stitching $K_z$ (\ie, the mapper of the filter function $z$) to $K_x$.
\end{example}

These $\LHD$ vectors have some interesting properties.
For instance, we know that $\LHD_0 = 0$ when stitching the mapper $K_x$  to $K_z$, since including more filter functions will not split any path-connected components; otherwise, they would have been represented by a cover element of the filter function $z$ in the univariate mapper $K_z$ already.
Additionally,  $\LHD$ can be shown to be always nonnegative.

\subsection{Local Relative Euler Characteristic}
\label{sec:LREC}

The local homology (LH) vectors can be summarized with the local relative Euler characteristic ($\LREC$) by computing the Euler characteristic restricted to each interval set, that is, the alternating sums of each homology class vector. 
Let $\chi(K)$ denote the Euler characteristic of a simplicial complex $K$. 

\begin{definition}[Local Relative Euler Characteristic]
 Given a pair of mappers $K_f:=M(f, \Ucal)$ and $K_g = M(g, \Vcal)$, we define $\LREC$ as a vector that captures the extent to which $K_g$ effects the Euler characteristic within each interval set $U_i \in \Ucal$ by stitching $K_g$ with $K_f$. 
 That is, we define Euler characteristic vectors for $K_f$ and $K_{(f,g)}$ and compute their difference,
 \begin{align}
 &   \chi^f = \left( \chi(K_f(U_1)), \chi(K_f(U_2)), \dots, \chi(K_f(U_k))\right),\\
 & \chi^{(f,g)} = \left( \chi(K_{(f, g)}(U_1)), \chi(K_{(f, g)}(U_2)), \dots, \chi(K_{(f, g)}(U_k)) \right), \\
& \mathtt{LREC} \left(K_f, K_{(f,g)}\right) = \chi^{(f,g)} - \chi^f.
 \end{align}
\end{definition}

\noindent Example~\ref{lrec} below demonstrates the $\LREC$ calculation on the cylinder from Example~\ref{lhd}. 

\begin{example}[$\LREC$ on a cylinder]
\label{lrec}
$$\LREC\left(K_x, K_{(x,z)}\right) = \begin{pmatrix}1 - 1 \\ 2-2 \\ 1-1 \end{pmatrix} =  \begin{pmatrix} 0 \\ 0\\ 0  \end{pmatrix}$$
$$\LREC\left(K_z, K_{(x,z)}\right) = \begin{pmatrix}0 - 1 \\ 0-1 \\ 0-1 \end{pmatrix} =  \begin{pmatrix} -1 \\ -1 \\ -1 \end{pmatrix}$$

As before, we illustrate this computation on a single interval set.
First, consider the first interval set $U_1$ of the $x$ projection.
$K_x$ has one vertex in $U_1$ (based on its corresponding interior subcomplex), so $\chi(K_{x}(U_1)) = 1$. 
$K_{(x,z)}$ restricted to the interval $U_1$ contains a line of 3 vertices with 2 edges, so $\chi(K_{(x, z)}(U_1)) = 3 -2 = 1$. 
Thus, for the first entry in $\LREC$, we have
$$\chi(K_{(x, z)}(U_1)) - \chi(K_{x}(U_1)) = 1 - 1 = 0.$$
In contrast, the first interval set of $K_z$ contains 1 vertex, that is $\chi(K_{z}(V_1)) = 1$; whereas $K_{(x,z)}$ within the same interval set contains a loop, that is, $\chi(K_{(x, z)}(V_1)) = 0$. Thus, for $V_1 \in \Vcal$, 
$$\chi(K_{(x, z)}(V_1)) - \chi(K_{z}(V_1)) = 0 - 1 = -1.$$

However, the $\LREC$ being negative does not necessarily mean that the topological information has decreased. 
The Euler characteristic does not scale with the expected information change in an intuitive way, and so $\LREC$ may be a less interpretable measurement in comparison with $\LHD$.
\end{example}

\subsection{Localized Entropy Differences}
\label{sec:LED}

Finally, the localized entropy difference (LED) compares the graph entropy for each interval set between two mapper graphs. 
Graph entropy was introduced~\cite{Rashevsky1955, Trucco1956} to measure the structural complexity of a graph, where it was originally referred to as the \emph{topological information content}~\cite{Rashevsky1955} of a graph. 
This is fitting for our purpose since we are interested in measuring the change of topological information content from a univariate mapper graph to a bivariate mapper graph.  
A number of graph entropy measures exist, see the survey by Dehmer and Mowshowitz~\cite{DehmerMowshowitz2011}.
We generalize and implement two of them in our visualization tool, although our framework is easily extendable to other entropy measures. 
For the remainder of this section, mapper graphs $G_f$ and $G_{(f,g)}$ represent the 1-skeletons of mappers $K_f := M(f, \Ucal)$ and $K_{(f,g)} := M((f,g), \Ucal \times \Vcal)$, respectively.

\begin{definition}[Localized Entropy Difference]
 Given a pair of mapper graphs $G_f$ and $G_{(f,g)}$, we define $\LED$ as a vector that captures the extent to which $G_g$ affects the graph entropy   within each interval set $U_i \in \Ucal$. 
 That is, we define localized entropy (LE) vectors for $G_f$ and $G_{(f,g)}$ and compute their difference,
 \begin{align}
 &   H^f = \left( H(G_f(U_1)), H(G_f(U_2)), \dots, H(G_f(U_k))\right),\\
 & H^{(f,g)} = \left( H(G_{(f, g)}(U_1)), H(G_{(f, g)}(U_2)), \dots, H(G_{(f, g)}(U_k)) \right), \\
& \mathtt{LED}\left(G_f, G_{\left(f,g\right)}\right) = H^{(f,g)} - H^f.
 \end{align}
\end{definition}
Here, $G_f(U_i)$ and $G_{\left(f,g\right)}(U_i)$ can be either interior or boundary mapper subgraphs. 
$H$ represents a certain notion of entropy. 
We introduce two types of LEDs, based on distance matrices and adjacency matrices, respectively. 

\para{Graph entropy based on the distance matrix.}
For an unweighted connected graph $G$, Bonchev and Trinajsti\'{c}~\cite{BonchevTrinajstic1977} introduced an entropy measure based on its graph distance matrix $D$. 
This measure originates from a notion in information theory called the \emph{information content} (\ie, \emph{Shannon entropy}) of a system. 

Assume a system $S$ contains $N$ elements. 
Consider all the $N$ elements are partitioned into $m$ groups, and $N_i$ is the number of elements in the $i$-th group.
We define the probability $p_i$ for a randomly selected element of $S$ to be found in the $i$-th group as $p_i = \frac{N_i}{N}$. 
Specifically, we work with the \emph{mean information content} of one element of the system, defined by Shannon's relation 
\begin{align}
\label{eq:shannon}
H(S) = - \Sigma_{i=1}^{m} p_i \log{p_i}.  
\end{align}

To apply~\cref{eq:shannon} to the setting of a graph $G$ with $N$ vertices and introduce an information measure on its distance matrix $D$, we consider all $N^2$ matrix elements in $D$ as elements of a system. 
Since $G$ is an unweighted graph, the distance of a value $i$ (where $0 \leq i \leq N-1$) appears $2 k_i$ times in $D$. 
Let $m$ be the highest value of $i$, which equals the diameter of the graph.  
Then a total of $N^2$ matrix elements are partitioned into $m\text{+}1$ groups, which correspond to distances valued at $\{0, 1, \dots, m\}$ respectively, where the  value of $0$ shows up $N$ times.
We associate each group with the probability for a randomly chosen distance to be in the $i$-th group. 
That is,  $p_i = \frac{2k_i}{N^2}$ and $p_0 = \frac{N}{N^2} = \frac{1}{N}$.
Applying~\cref{eq:shannon} to $m\text{+}1$ groups, the \emph{mean information on distances} of a graph $G$ is defined as~\cite[Eq.~(8)]{BonchevTrinajstic1977} 
\begin{align}
\label{eq:mean-info-onecpnt}
  H_D(G) = - \frac{1}{N}\log(\frac{1}{N}) - \Sigma_{i=1}^{m}\frac{2k_i}{N^2}\log(\frac{2k_i}{N^2}).
\end{align}
In this paper, we generalize \cref{eq:mean-info-onecpnt} to graphs with multiple connected components by considering $\infty$ as an additional possible value in the distance matrix $D$.
That is, we define $p_\infty = \frac{2 k_\infty}{N^2}$, where $2k_\infty$ is the number of times a value $\infty$ appears in $D$. 
Therefore, for a graph $G$ whose matrix $D$ might contain $\infty$ (that is, $G$ contains multiple connected components), we apply \cref{eq:mean-info} with $m\text{+}2$ groups to define a \emph{mean entropy on distance}, 
\begin{align}
\label{eq:mean-info}
H_D(G) = - \frac{1}{N}\log(\frac{1}{N}) - \Sigma_{i=1}^{m}\frac{2k_i}{N^2}\log(\frac{2k_i}{N^2}) - \frac{2k_\infty}{N^2}\log(\frac{2k_\infty}{N^2}).
\end{align}
Based on \cref{eq:mean-info}, we define \emph{$\LED$ based on distances} as 
\begin{align} 
\mathtt{LED}_D \left(G_f, G_{(f,g)}\right) = H_D^{(f,g)} - H_D^f.
\end{align} 
Intuitively speaking, $H_D(G)$ captures the information on the distribution of distances in the graph; it has been shown to be useful experimentally in studying the branching of graphs having different numbers of vertices~\cite{BonchevTrinajstic1977}.
Therefore, $\mathtt{LED}_D$ quantifies changes in branching structures moving from a univariate to a bivariate mapper graph.

\para{Graph entropy based on the adjacency matrix.}
Mackenzie~\cite{Mackenzie1966} proposed an entropy based on {an adjacency} matrix, which was employed by Sen \etal~\cite{SenChuParhi2019} to study brain networks. 
For a weighted graph $G$, let $w_{ij}$ be the weight of edge $e_{ij}$ between vertices $v_i$ and $v_j$. 
Let $W = \Sigma_{e_{ij} \in E} (w_{ij})$ be the total edge weight. 
The probability of correlation between $v_i$ and $v_j$ is defined~\cite[Eq.~5]{SenChuParhi2019} as
$$q_{ij} = 
\begin{cases}
\frac{w_{ij}}{W}& \text{when } i \neq j\text{, and }\\
0& \text{when } i=j.
\end{cases}$$
The \emph{mean entropy on adjacency} is then defined as~\cite[Eq.~6]{SenChuParhi2019} 
\begin{align}
\label{eq:mean-adjacency}
  H_A(G) = - \Sigma_{e_{ij}\in E} \left( q_{ij} \log(q_{ij}) \right).
\end{align}
We extend \cref{eq:mean-adjacency} to handle mapper subgraphs that are not necessarily connected by considering $q_{ij} = 0$ when $v_i$ and $v_j$ belong to different connected components of $G$.
Based on \cref{eq:mean-adjacency}, we define \emph{$\LED$ based on adjacencies} as  
\begin{align} 
  \mathtt{LED}_A \left(G_f, G_{(f,g)}\right) = H_A^{(f,g)} - H_A^f.
\end{align} 
Intuitively, $H_A(G)$ captures the centrality property of a graph: it varies inversely with respect to the structural centrality of $G$, that is, $H_A(G)$ increases as the graph becomes decentralized~\cite{Mackenzie1966}. 
It can be used to compare a pair of graphs of different sizes, where a graph with a smaller entropy indicates more centrality thus less randomness~\cite{SenChuParhi2019} in its structure.   Therefore, $\mathtt{LED}_A$ quantifies changes in centrality moving from a univariate to a bivariate mapper graph.

\para{Remarks.} 
Finally, we note that a number of entropy measures are defined for simplicial complexes (e.g.,~\cite{DantchevIvrissimtzis2017}), which may be applicable to mappers (not just mapper graphs).
This is left for future work.

\section{Visualizing Topological Gains}
\label{sec:results}

We provide a tool that visualizes topological gains during the stitching process. 
We experiment with two synthetic 2- or 3-dimensional point cloud datasets together with four classic datasets in machine learning, the Boston Housing dataset, the Iris dataset, the Breast Cancer dataset, and the Wine Quality dataset, some of which are available via the UCI Machine Learning Repository~\cite{DuaGraff2017}.
We also explore two real-world datasets, a phenomics dataset referred to as the KS/NE dataset and a breast cancer dataset referred to as the NKI dataset.  
For each dataset $\Xspace$, we compare mapper graphs $G_f$ and $G_{(f,g)}$ constructed based on a pair of variables $f, g: \Xspace \to  \Rspace$. We implement localized homological differences in dimensions  0 and 1 (denoted as $\LHD_0$ and $\LHD_1$), as well as localized entropy differences based on distances ($\LED_D$) and adjacencies ($\LED_A$), respectively. 

\para{Implementation details.}
We implement the tool using HTML/CSS/JavaScript stack with \textit{D3.js} and \textit{JQuery} libraries. 
It interfaces with a Python backend using a \textit{Flask}-based server. 
The tool is an extension of \textit{Mapper Interactive}~\cite{ZhouChalapathiRathore2021}, which is an extendable and interactive toolbox for the visual exploration of high-dimensional data using the mapper algorithm. 
In particular, \textit{Mapper Interactive} uses an accelerated modification  of \textit{KeplerMapper}~\cite{VeenSaulEargle2019} to compute mapper graphs in a scalable way.

\subsection{Visualization Interface}
\begin{figure}[!ht]
\centering
\includegraphics[width=\columnwidth]{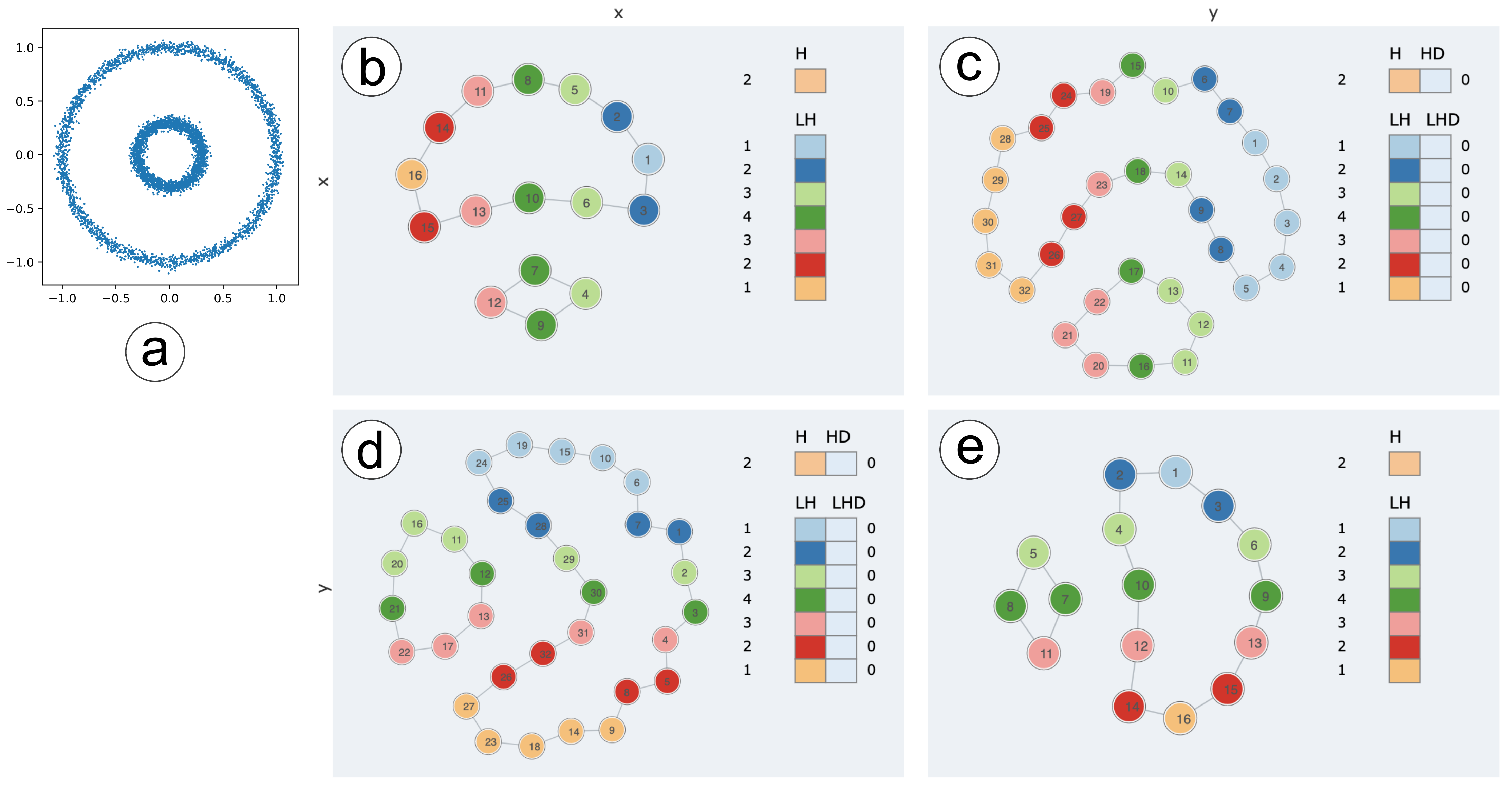}
\vspace{-2mm}
\caption{\emph{Two Circles} dataset: the mapper graph matrix with $\LHD_0$. Graph nodes are colored by interval indices.}
\label{fig:2-circles-lhd}
\vspace{-2mm}
\end{figure}

We begin with an example of a 2-dimensional point cloud $\Xspace = \{(x_i,y_i)\}$ containing two nested circles in~\autoref{fig:2-circles-lhd}a to illustrate our main visualization interface.  
The mapper graph parameters are $n = 7$, $p = 5\%$. 
The two filter functions are chosen to be the $x$- and $y$-coordinates of the points, $x, y: \Xspace \to \Rspace$.

The main display is in the form of a mapper graph matrix, shown  in~\autoref{fig:2-circles-lhd} . 
As illustrated in~\autoref{fig:2-circles-lhd}, we construct univariate  mapper graphs $G_x$ in (b) and $G_y$ in (e) for variables $x$ and $y$, respectively, that are placed along the diagonal of the mapper graph matrix. 
We construct bivariate mapper graphs $G_{(x,y)}$ in (c) and (d), respectively, which are shown off-diagonal. 
The mapper graphs are drawn with force-directed layouts. 
Nodes are colored by the index of intervals, where indices $1$ to $7$ correspond to the color light blue, dark blue, light green, dark green, pink, red, and orange, respectively.

For each mapper graph, we report its associated 0-dimensional local homology (LH) vectors w.r.t.~the interior mapper subgraphs. 
For instance, in~\autoref{fig:2-circles-lhd}e, the LH vector $\beta^{y}_0 = (1,2,3,4,3,2,1)$ captures the number of connected components for the univariate mapper graph $G_y$ of variable $y$. 
The LH vector $\beta_0^{(x,y)} = (1,2,3,4,3,2,1)$ in~\autoref{fig:2-circles-lhd}d similarly captures the distribution of connected components for the bivariate mapper graph $G_{(x,y)}$ along the intervals for $y$.  
For each bivariate mapper graph $G_{(x,y)}$, we also report its localized homological difference ($\LHD_0$) w.r.t.~$G_x$ and $G_y$, respectively.
For instance, $\LHD_0(G_x, G_{x,y}) = 0$ (\autoref{fig:2-circles-lhd}c) and $\LHD_0(G_y, G_{x,y}) = 0$ (\autoref{fig:2-circles-lhd}d, due to symmetry).  

In general, within a mapper graph matrix, univariate and bivariate mapper graphs are placed on and off the diagonal, respectively.  
For each mapper graph, the graph nodes are colored by the  intervals they belong to: they are either colored by interval indices or by the value of a measure attached to each interval. 
The rectangle bars on the right of each mapper graph demonstrate the vector of either $\LH$ or $\LE$ restricted to the interval sets. 
They are either colored by interval indices (\autoref{fig:2-circles-lhd}), or by a continuous colormap associated with the values of a chosen measure.  
For the bivariate mapper graphs, we also display $\LHD$/$\LED$ between the bivariate and univariate mapper graphs. 
 Such differences are computed by subtracting the values of $\LH$ or $\LE$ in a univariate mapper graph from the values in a  bivariate mapper graph on the same row.

By looking at $\LHD$ or $\LED$, we would know how much topological information is gained during the stitching process. In addition, by comparing the $\LHD$ or $\LED$ between two bivariate mapper graphs, we could identify the variables with high $\LHD$/$\LED$ values. 
 Such a variable is considered to be more important than the other variables in terms of extracting topological information of a  given point cloud. More detailed explanations of such comparisons will be provided in the examples below.

\subsection{Cylinder}

\begin{figure*}[!ht]
\centering
\includegraphics[width=0.98\columnwidth]{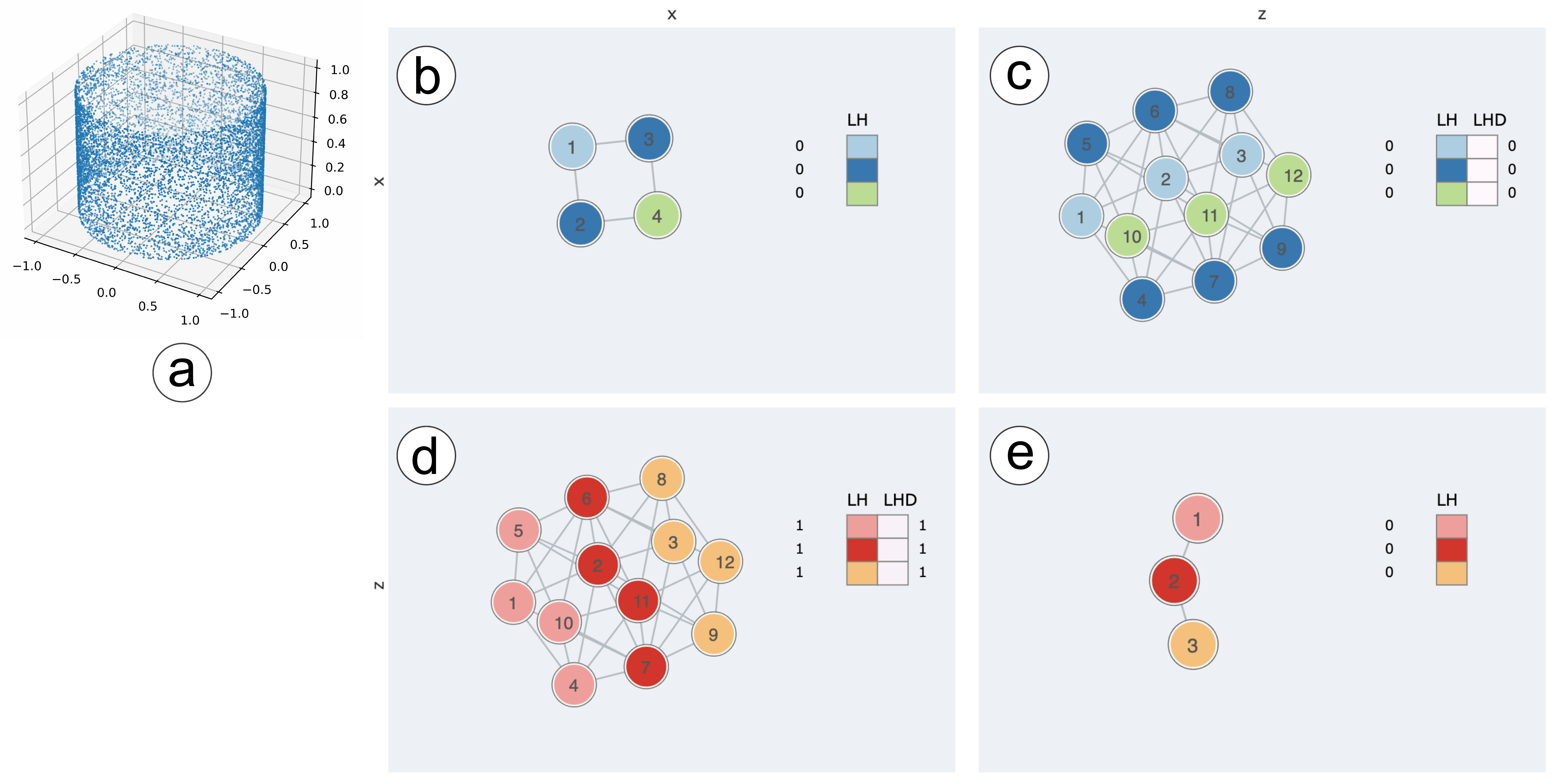}
\vspace{-2mm}
\caption{\emph{Cylinder} dataset: the mapper graph matrix with $\LHD_1$. Graph nodes are colored by interval indices.}
\label{fig:cylinder-lhd}
\vspace{-2mm}
\end{figure*}

Our first example is to reproduce the result of Example~\ref{lhd} by generating a 3-dimensional cylinder point cloud $\Xspace = \{(x_i, y_i, z_i)\}$, where the $x$-, $y$-, and $z$-coordinates correspond to the three filter functions, respectively. 
The point cloud is shown in~\autoref{fig:cylinder-lhd}a.
The mapper graph parameters are $n=3$, $p=15\%$. 
The two filter functions are chosen to be the $x$- and $z$- coordinates of the points, $x, z: \Xspace \to \Rspace$.
The two univariate mapper graphs $G_x$ in~\autoref{fig:cylinder-lhd}b and $G_z$ in~\autoref{fig:cylinder-lhd}e for variables $x$ and $z$, respectively, are placed along the diagonal of \autoref{fig:cylinder-lhd}.
The bivariate mapper graphs $G_{(x,z)}$ are off diagonal in~\autoref{fig:cylinder-lhd}c and~\autoref{fig:cylinder-lhd}d, respectively.

For each mapper graph, we report its associated 1-dimensional local homology (LH) vectors w.r.t.~the interior mapper subgraphs. The results confirm our previous computation in \autoref{sec:LHD} that $\LHD_1$ increases when stitching $G_x$ to $G_z$, while $\LHD_1$ does not increase when stitching $G_z$ to $G_x$ (see~\autoref{fig:cylinder-lhd}d and~\autoref{fig:cylinder-lhd}c).

\subsection{Sphere}

\begin{figure}[!ht]
\centering
\includegraphics[width=\columnwidth]{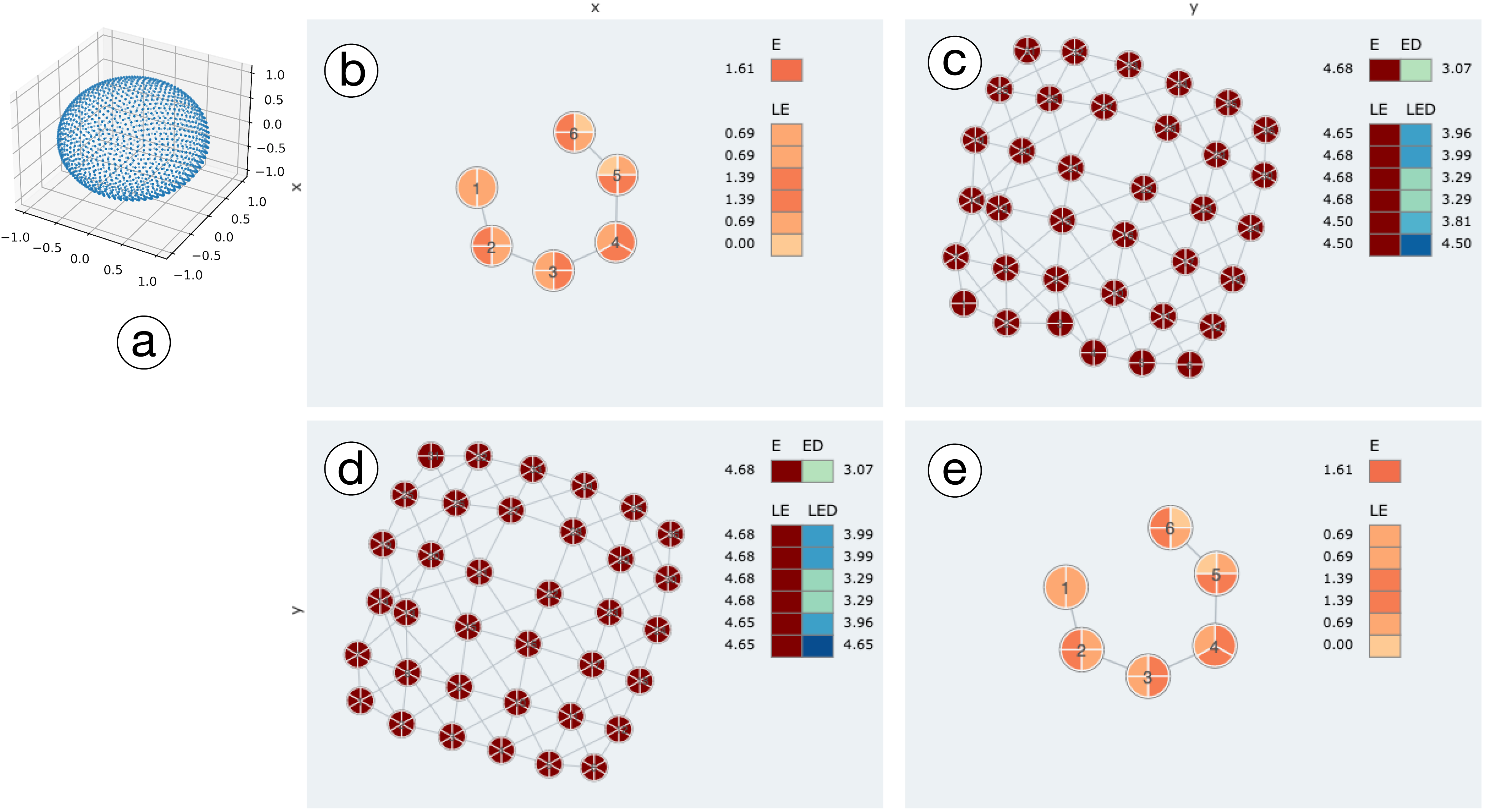}
\vspace{-3mm}
\caption{\emph{Sphere} dataset: the mapper graph matrix with $\LED_A$. Graph nodes are colored by $LE$ measures restricted to the interval sets.}
\label{fig:sphere-am}
\vspace{-3mm}
\end{figure}

Our second example is a 3-dimensional point cloud sampled from the surface of a sphere (\autoref{fig:sphere-am}a). 
We again choose 2 of the 3 dimensions ($x$ and $y$) to compute the mapper graphs $G_x$ and $G_y$, as well as the boundary mapper subgraphs. 
The mapper parameters are $n = 6$, $p = 15\%$. 
In this example, we observe the information content, quantified by localized entropy difference (LED) based on adjacencies ($\LED_A$), increases in the bivariate mapper graphs, especially for the intervals capturing the top and bottom of the sphere. 
For instance, the localized entropy (LE) vectors $H^x$, $H^{(x,y)}$, and their difference are shown in \autoref{fig:sphere-am}b and \autoref{fig:sphere-am}c, respectively, where  
\begin{align*}
H^x & = \begin{pmatrix} 0.00, & 0.69, & 1.39, & 1.39, & 0.69, & 0.69 \end{pmatrix},\\
H^{(x,y)} & = \begin{pmatrix} 4.50, & 4.50, & 4.68, & 4.68, & 4.68, & 4.65 \end{pmatrix},\\
\LED_A(H^x, H^{(x,y)}) & = \begin{pmatrix} 4.50, & 3.81, & 3.29, & 3.29, & 3.99, & 3.96 \end{pmatrix}. 
\end{align*}
In this example, LHD does not give much information in dimension 0 since $\beta^x_{0} = \beta^{(x,y)}_{0} = (1,1,1,1,1,1)$, hence giving $\LHD_0 = 0$.  

\begin{figure*}[!ht]
\centering
\includegraphics[width=\columnwidth]{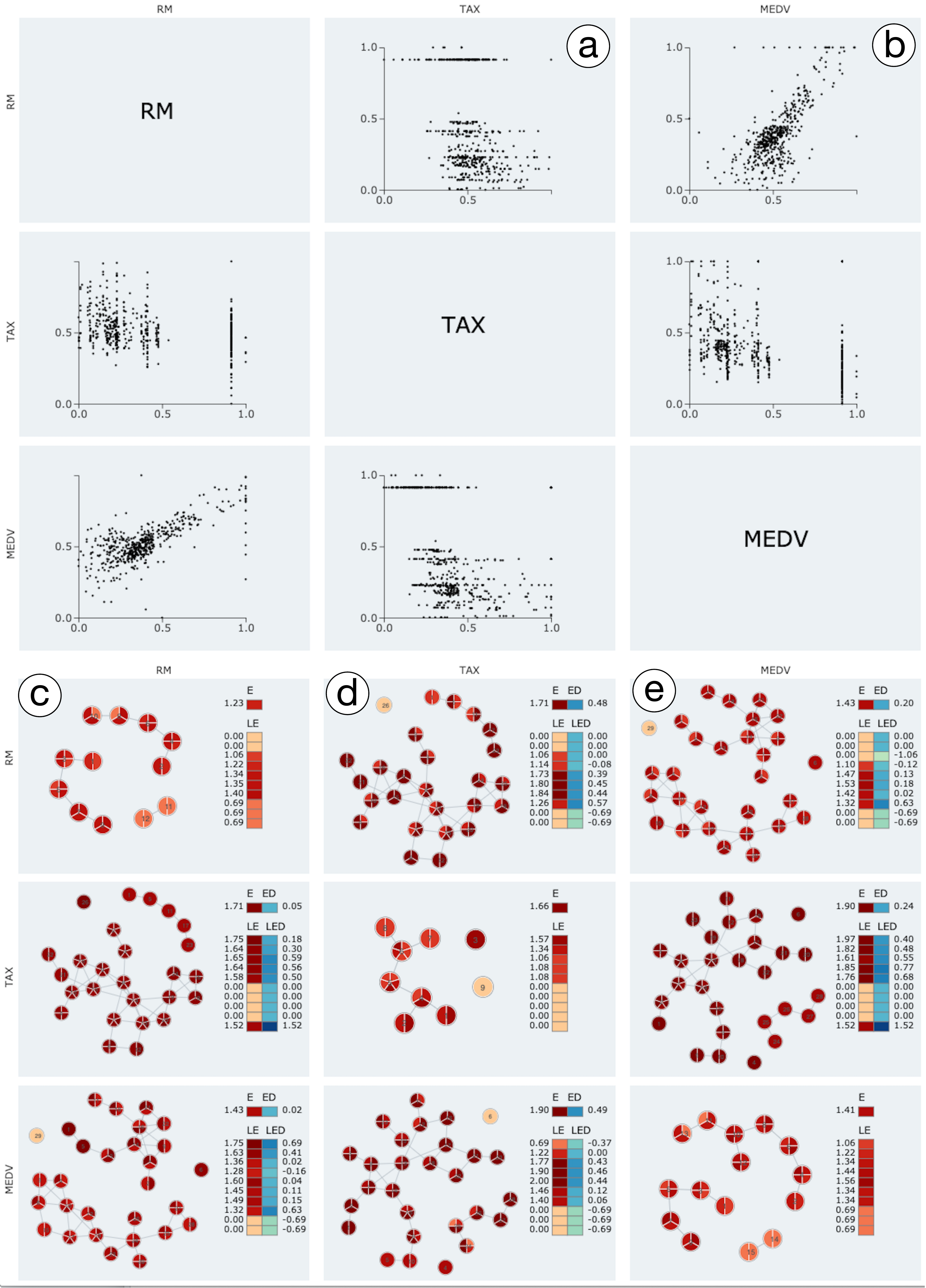}
\vspace{-2mm}
\caption{\emph{Boston Housing} dataset. Top: the scatter plot matrix. Bottom: the mapper graph matrix with $\LED_D$; graph nodes are colored by $LE$ restricted to interval sets.}
\label{fig:boston-dm}
\vspace{-2mm}
\end{figure*}

\subsection{Boston Housing Dataset}
Our third example is the classic Boston Housing dataset~\cite{HarrisonRubinfeld1978}, which contains housing information in the Boston area collected by the U.S.~Census Service. 
It contains 14 attributes per data point, including \emph{CRIM} (per capita crime rate by town), \emph{RAD} (index of accessibility to radial highways), and \emph{ZN} (proportion of residential land zoned for lots over 25,000 sq.~ft.). 
We chose three attributes as variables to compute the mapper graphs: 
\emph{RM}, which is the average number of rooms per dwelling, 
\emph{TAX}, which is the full-value property-tax rate per 10,000 dollars, and \emph{MEDV}, which is the median value of owner-occupied homes in 1000's of dollars, using mapper parameters $n = 10$, $p=15\%$. 
We compute LED based on distances, denoted as $\LED_D$, for boundary mapper subgraphs.
For instance, for variables \emph{RM} and \emph{TAX} (\autoref{fig:boston-dm}c, \autoref{fig:boston-dm}d),   
\begin{align*}
H^{RM} & = (0.69, 0.69, 0.69, 1.4,1.35, 1.34, 1.22, 1.06, 0, 0),\\
H^{(RM, TAX)} & = (0, 0, 1.26,1.84, 1.73, 1.14, 1.06, 0,0 ),\\
\LED_D(H^{RM}, H^{(RM, TAX)}) &= (-0.69, -0.69, 0.57, 0.44, 0.45, 0.39, -0.08, 0, 0, 0). 
\end{align*}
The mapper graph matrix complements the scatter plot matrix in \autoref{fig:boston-dm}. 
In particular, we observe globally that stitching the mapper graph $G_{TAX}$ to $G_{RM}$ has a higher global LED ($0.48$) in comparison to stitching $G_{MEDV}$ to $G_{RM}$ ($0.20$), which is aligned with the observation that \emph{RM} vs.~\emph{TAX} are less correlated than \emph{RM} vs.~\emph{MEDV} (see~the scatter plots in \autoref{fig:boston-dm}a and \autoref{fig:boston-dm}b).
Understanding the relation between topological gains and correlations among variables will be an interesting future direction.

\subsection{Iris Dataset}
\begin{figure}[!ht]
\centering
\includegraphics[width=\columnwidth]{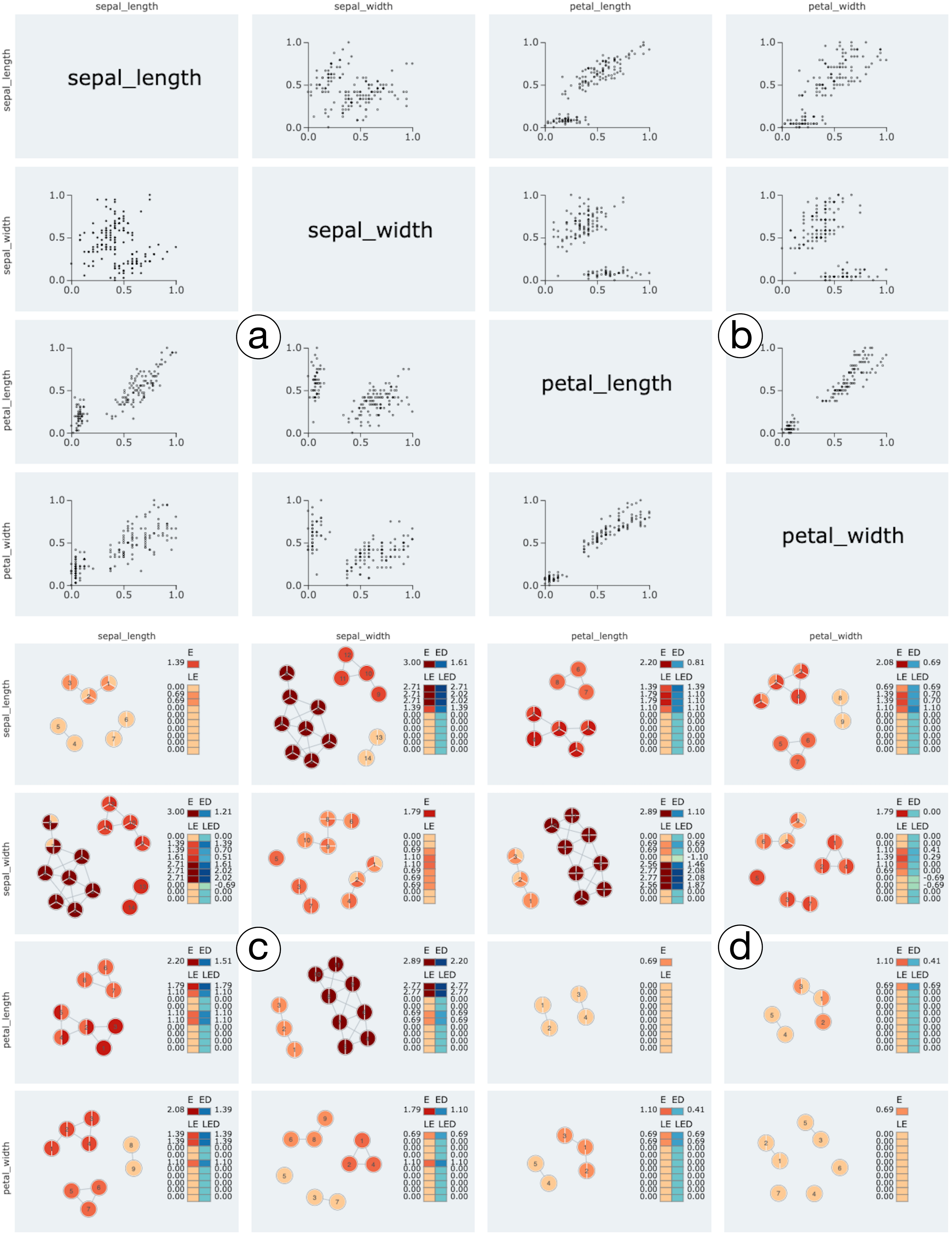}
\vspace{-2mm}
\caption{\emph{Iris} dataset. Top: the scatter plot matrix. Bottom: the mapper graph matrix with $\LED_A$; graph nodes are colored by $LE$ restricted to interval sets.}
\label{fig:iris-am}
\vspace{-2mm}
\end{figure}

Our fourth example is Fisher's Iris dataset~\cite{Fisher1936}, another classic dataset in machine learning. 
This dataset contains four attributes including the \emph{sepal length}, \emph{sepal width}, \emph{petal length}, and \emph{petal width} of each iris plant. 
We use all four attributes to compute the mapper graph matrix, with mapper parameters $n = 10$, $p = 30\%$. 
As shown in~\autoref{fig:iris-am}, we combine both the scatter plot matrix and the mapper graph matrix.  
We observe that stitching the mapper graph associated with \emph{sepal width} to \emph{petal length} has a higher global $\LED_A$ (\autoref{fig:iris-am}c, $2.20$) than stitching \emph{petal width} to \emph{petal length} (\autoref{fig:iris-am}d, $0.41$). 
Such a topological gain is also observed locally for boundary subgraphs. 
At the same time, \emph{petal length} is shown to be more correlated to \emph{petal width} than with \emph{sepal width} (see~\autoref{fig:iris-am}b and \autoref{fig:iris-am}a).

\begin{figure*}[!ht]
\centering
\includegraphics[width=\columnwidth]{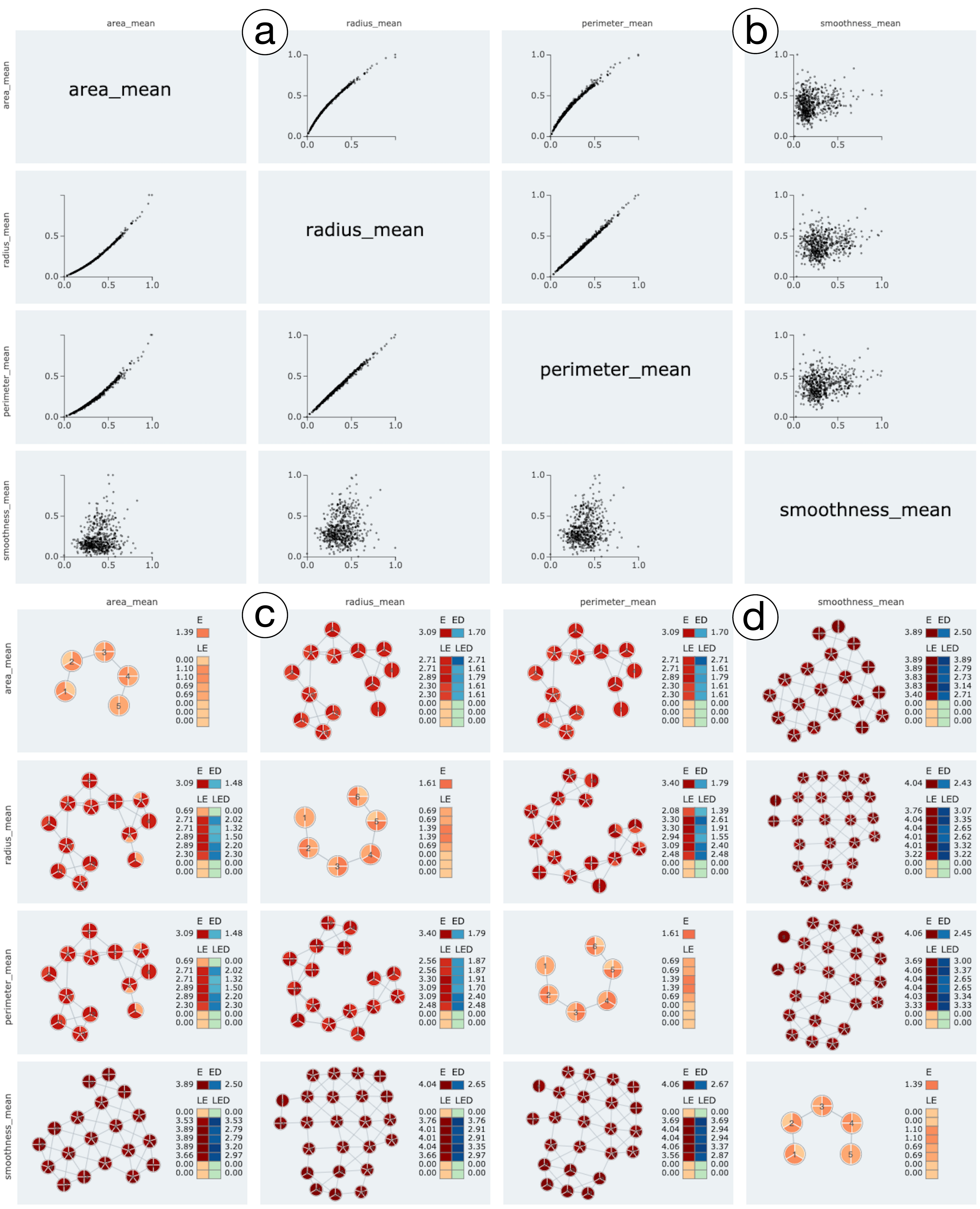}
\vspace{-2mm}
\caption{\emph{Breast Cancer} dataset. Top: the scatter plot matrix. Bottom: the mapper graph matrix with $\LED_A$; graph nodes are colored by $LE$ restricted to interval sets.}
\label{fig:breast-cancer-am}
\vspace{-2mm}
\end{figure*}

\subsection{Breast Cancer Wisconsin (Diagnostic) Dataset}
Our fifth dataset describes characteristics of the cell nuclei present in the images of breast masses~\cite{MangasarianStreetWolberg1995,StreetWolbergMangasarian1993}. 
We choose four variables from among ten real-valued features computed for each cell nucleus: \emph{area mean} (mean area of the tumor), \emph{radius mean} (mean of distances from the center to points on the perimeter), \emph{parameter mean} (mean size of the core tumor), and \emph{smoothness mean} (mean of local variation in radius lengths). 
We compute univariate and bivariate mapper graphs using boundary subgraphs, with parameters $n = 8$, $p = 20\%$.  
As shown in the scatter plot matrix (\autoref{fig:breast-cancer-am}),  \emph{area mean} and \emph{radius mean} are highly correlated,  but \emph{area mean} and \emph{smoothness mean} are not. 
We observe that stitching the mapper graph of the \emph{smoothness mean} to that of the \emph{area  mean} achieves a higher global and local $\LED_A$ than stitching \emph{radius mean} with \emph{area mean} (see~\autoref{fig:breast-cancer-am}d and \autoref{fig:breast-cancer-am}c).

\subsection{Wine Quality Dataset}

\begin{figure*}[!ht]
\centering
\includegraphics[width=\columnwidth]{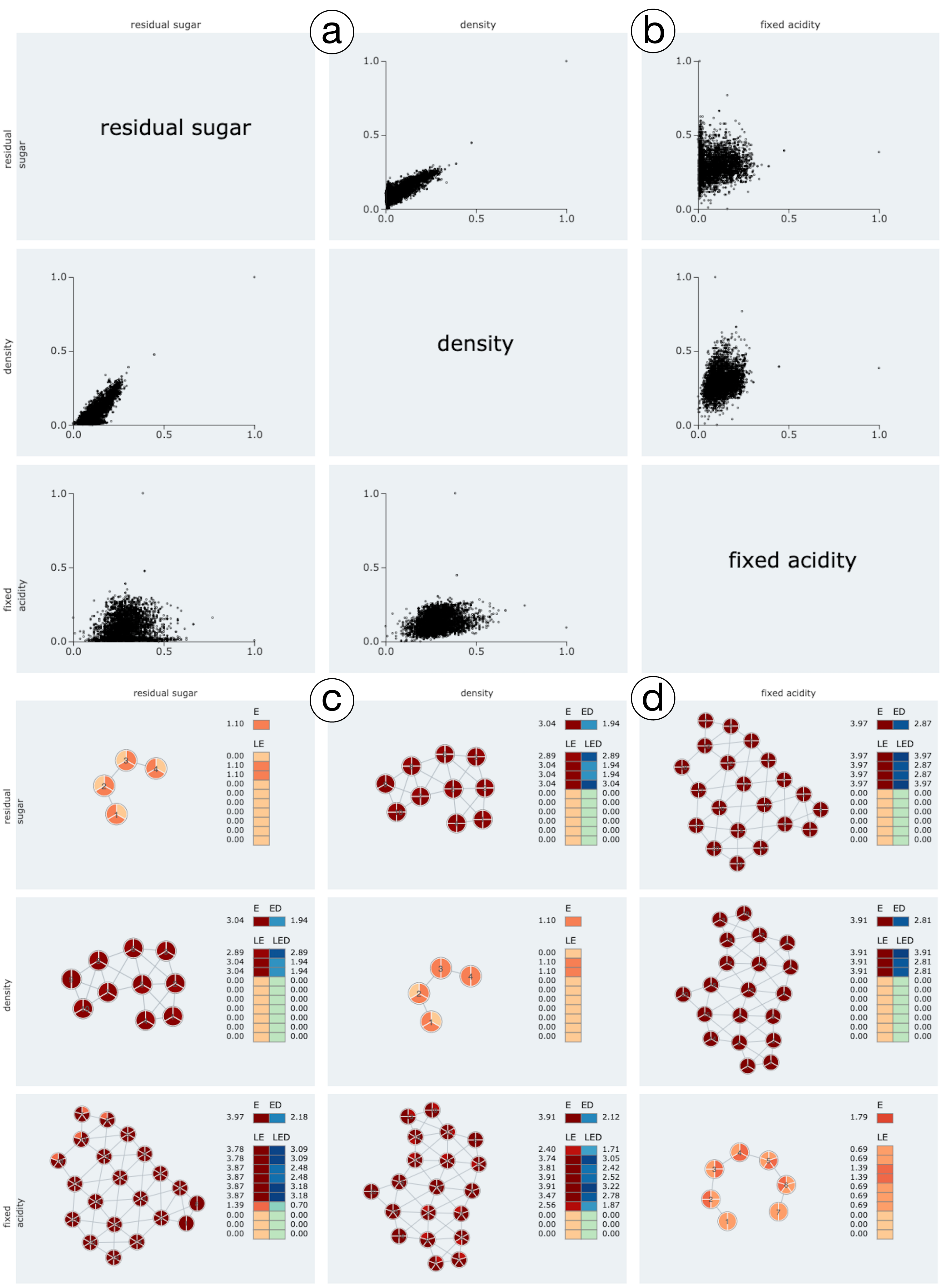}
\vspace{-2mm}
\caption{\emph{Wine Quality} dataset. Top: the scatter plot matrix. Bottom: the mapper graph matrix with $\LED_A$; graph nodes are colored by $LE$ restricted to interval sets.}
\label{fig:wine-quality-am}
\vspace{-2mm}
\end{figure*}

Our sixth dataset is the wine quality dataset, which is another classic machine learning dataset, and gives 11 variables describing wine quality based on physicochemical tests~\cite{CerdeiraAlmeidaMatos2009}. 
We use 3 of these variables for our analysis: \emph{residual sugar}, \emph{density}, and \emph{fixed acidity}. 
The scatter plot matrix shows that \emph{residual sugar} and \emph{density} are highly correlated (\autoref{fig:wine-quality-am}a), but \emph{residual sugar} and \emph{fixed acidity} are not (\autoref{fig:wine-quality-am}b). 
Complementarily, we see that stitching the mapper graph of \emph{fixed acidity} to the mapper graph of \emph{residual sugar} gives rise to higher $\LED_A$ globally and locally than stitching \emph{density} with \emph{residual sugar} (see~\autoref{fig:wine-quality-am}d and \autoref{fig:wine-quality-am}c).

\subsection{KS/NE dataset}
We also explore a real-world phenomics dataset referred to as the KS/NE dataset and was first studied by Kamruzzaman~\etal~\cite{KamruzzamanKalyanaramanKrishnamoorthy2019}.
It records daily measurements of maize plants that were cultivated in Kansas (KS) and Nebraska (NE). 
The columns consist of the genotype of each plant, the growth rate of each plant (\emph{growth\_rate}), a time measurement describing the days after planting (\emph{DAP}), and 10 environmental variables including \emph{humidity}, \emph{temperature}, \emph{rainfall}, \emph{solar radiation}, etc.
There are 400 rows, with each row corresponding to the daily record of a plant. We construct a 1D point cloud using \emph{growth\_rate}, and choose the variables \emph{DAP} and \emph{humidity} to compute the mapper graphs $G_{DAP}$, $G_{humidity}$, and $G_{(DAP, humidity)}$, as well as the boundary mapper subgraphs.
The mapper parameters are $n=10$, $p=46\%$.

\begin{figure}[!ht]
\centering
\includegraphics[width=0.9\columnwidth]{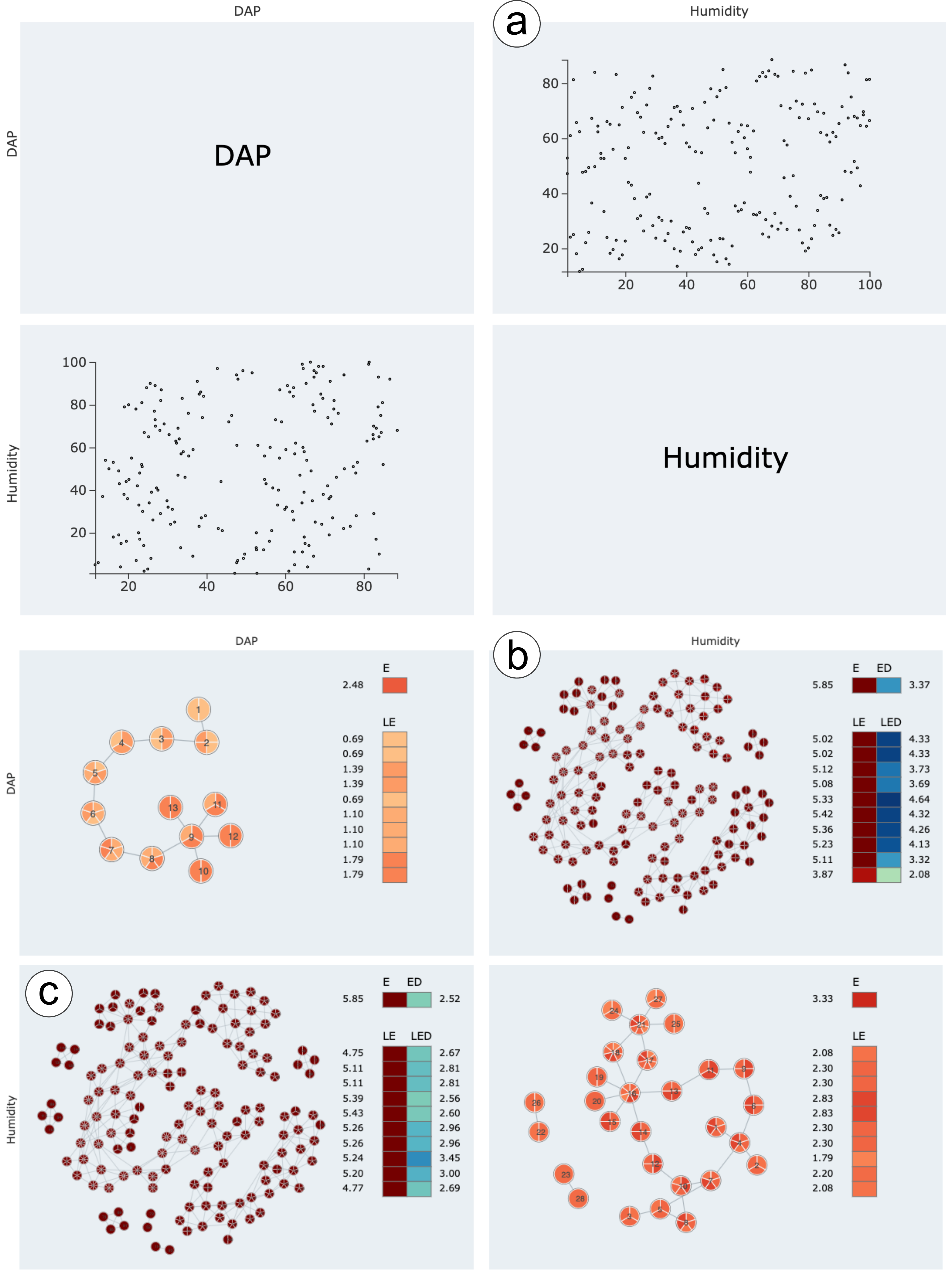}
\vspace{-2mm}
\caption{\emph{KS/NE} dataset: Top: the scatter plot matrix. Bottom: the mapper graph matrix with $\LED_A$; graph nodes are colored by $LE$ restricted to interval sets.}
\label{fig:ksne-am}
\vspace{-2mm}
\end{figure}

As shown in~\autoref{fig:ksne-am}, we observe that the bivariate mapper graphs have positive $\LED_A$ both globally and locally, indicating the bivariate mapper graphs have more topological gains than the univariate mapper graphs.
In particular, stitching $G_{humidity}$ to $G_{DAP}$ has a higher global $\LED_A$ compared with stitching $G_{DAP}$ to $G_{humidity}$ (see~\autoref{fig:ksne-am}b and~\autoref{fig:ksne-am}c), indicating that $humidity$ is a more important variable than $DAP$ for capturing the topological structure of the point cloud data.
The scatter plot matrix~\autoref{fig:ksne-am}a shows that the two variables are not correlated.

\subsection{NKI dataset}
We explore another real-world dataset, the breast cancer dataset that provides prognosis and gene expression information of patients.
This dataset is referred to as the NKI dataset~\cite{Vant-VeerDaiVan-De-Vijver2002}, which was previously studied by Lum~\etal~\cite{LumSinghLehman2013} and Zhou~\etal~\cite{ZhouChalapathiRathore2021} to identify subgroups in breast cancer patients.
It contains 272 rows, with each row corresponding to the information of a patient. The columns consist of 1554 gene expression levels, and variables of medical records or physiological measures such as \emph{event\_death} (whether a patient survived or not), \emph{survival\_time}, \emph{recurrence\_time}, etc.
We construct the point cloud using the 1500 mostly varying genes, and choose the variable \emph{event\_death} together with the infinity norm (\emph{$L_{\infty}$}) of the point cloud to compute the mapper graphs $G_{event\_death}$, $G_{L_{\infty}}$ and $G_{(event\_death, L_{\infty})}$, as well as the boundary mapper graphs.
The mapper parameters are $n=18$, $p=68\%$. 

\begin{figure}[!ht]
\centering
\includegraphics[width=0.9\columnwidth]{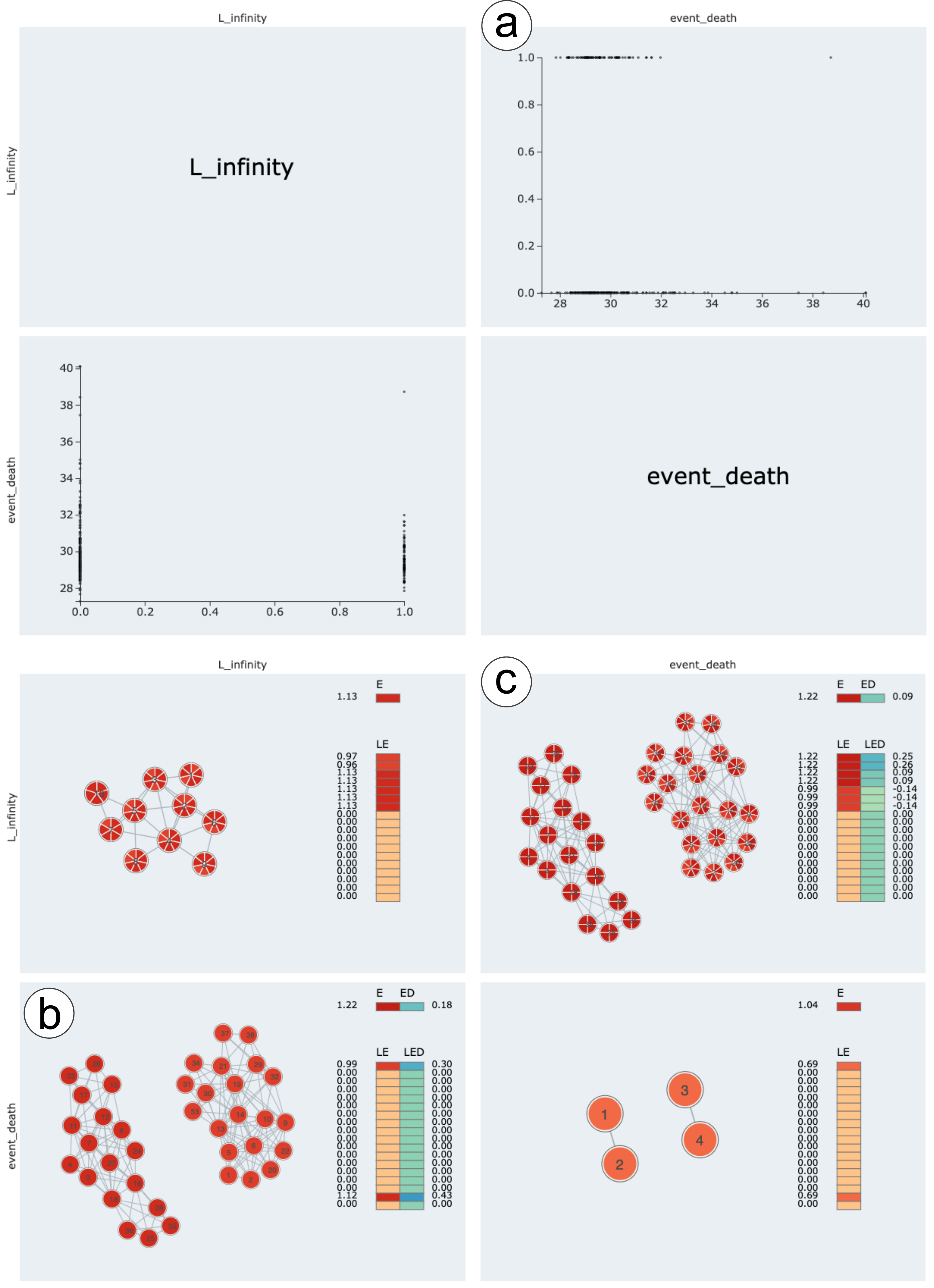}
\vspace{-2mm}
\caption{\emph{NKI} dataset: Top: the scatter plot matrix. Bottom: the mapper graph matrix with $\LED_D$; graph nodes are colored by $LE$ restricted to interval sets.}
\label{fig:nki-dm}
\vspace{-2mm}
\end{figure}

As shown in~\autoref{fig:nki-dm}, we observe that the bivariate mapper graphs have positive $\LED_D$ both globally and locally, indicating the bivariate mapper graphs have more topological gains than the univariate mapper graphs. In particular, stitching $G_{L_{\infty}}$ to $G_{event\_death}$ has a higher global $\LED_D$ compared with stitching $G_{event\_death}$ to $G_{L_{\infty}}$ (see~\autoref{fig:nki-dm}b and~\autoref{fig:nki-dm}c), indicating that $L_{\infty}$ is a more important variable than \emph{event\_death} for capturing the topological structure of the point cloud data.
The scatter plot matrix~\autoref{fig:nki-dm}a shows that the two variables are not correlated.

\section{Discussion}
\label{sec:discussion}

We study a method of stitching (composing) a pair of univariate mappers together into a bivariate mapper. 
By tracking the \texttt{STITCH} and \texttt{FIX} steps of the construction process, it is possible to quantify the relationship between filter functions.  
We further propose measures of topological gains that quantify the changes in topological content during the stitching process.   

With such measures in hand, we return to our topological analogues of the stepwise regression~\cite{Efroymson1960} and scatter plot matrix~\cite{ElmqvistDragicevicFekete2008}, which help to navigate topological relationships among multiple filter functions.
A method for \emph{stepwise stitching} would produce a mapper with optimal topological information by iteratively building a multivariate mapper from topologically independent filter functions. 
A \emph{topological scatter plot matrix} can reveal information about the filter functions such as topological dependencies and outliers by providing a visualization of the most information-rich filter functions.
Our visualization tool provides a playground for such types of future work. 
Furthermore, based on various datasets analyzed in~\autoref{sec:results}, we observe that stitching the mapper graphs of highly correlated variables typically gives rise to smaller changes in entropy ($\LED$) than stitching the mapper graphs of uncorrelated variables. Studying the relation between topological correlations (via mapper graphs) and statistical correlations will be an interesting future direction.


\begin{acknowledgement}
This research was partially supported by DOE DE-SC0021015, NSF DBI-1661375, DBI-1661348, and DMS-1819229. 
\end{acknowledgement}



\end{document}